\documentclass {article} % [sigconf]{acmart}
\usepackage{arxiv}
\usepackage{booktabs} % For formal tables
\usepackage{algorithm}% 
\usepackage{algpseudocode}% 
\usepackage{hyperref}
\usepackage{amsmath}
\usepackage{xcolor}
\usepackage{xspace}
\usepackage{soul}
\newcommand{\ignore}[1]{}
\usepackage[textsize=tiny]{todonotes}
\usepackage{graphicx}
\usepackage{boldline}
\usepackage[export]{adjustbox}
\usepackage[caption=false]{subfig}
\usepackage[font=small]{caption}
\usepackage{adjustbox}
\usepackage{enumerate}
\usepackage{amsfonts}
\usepackage{colortbl}
\usepackage{float}
\usepackage{rotating}
\usepackage{setspace}
\usepackage{gensymb}	% for degree symbol
\usepackage[official]{eurosym} % euro
\usepackage[flushleft]{threeparttable}
\usepackage{enumitem}
\usepackage{array}
\title{Ocean Wave Energy Converters Optimization: A Comprehensive Review on Research Directions}

\author{
  Danial Golbaz \\
  School of Civil Engineering\\
  College of Engineering\\
  The University of Tehran\\
  Iran \\
  \texttt{Dgolbaz@ut.ac.ir} \\
\And
  Rojin Asadi \\
  School of Civil Engineering\\
  College of Engineering\\
  The University of Tehran\\
  Iran \\
  \texttt{rojinasadi@ut.ac.ir} \\

\And

Erfan Amini\\
School of Civil Engineering\\
College of Engineering\\
  University of Tehran\\
  Tehran 13145, Iran\\
	 \texttt{erfan.amini@ut.ac.ir} \\
	 
\And
  Hossein Mehdipour \\
 School of Civil Engineering\\
  College of Engineering\\
  The University of Tehran\\
  Iran \\
  \texttt{hossein.mehdipour@ut.ac.ir} \\
  
  \And
  Mahdieh Nasiri \\
  School of Mechanical Engineering \\
  Iran University of Science and Technology\\
  Iran\\
  \texttt{mahdie\_nasiri@alumni.iust.ac.ir} \\
  
  \And
  Meysam Majidi Nezhad \\
  Department of Astronautics \\
Electrical and Energy Engineering (DIAEE)\\
	Sapienza University of Rome\\
	Italy\\
  \texttt{meysam.majidinezhad@uniroma1.it} \\
  
  \And
  Seyed Taghi Omid Naeeni \\
  Associate Professor in School of Civil Engineering\\
  College of Engineering\\
  The University of Tehran \\
  Iran\\
  \texttt{meysam.majidinezhad@uniroma1.it} \\

  \And
   Mehdi Neshat \\
  Optimization and Logistics Group\\
  School of Computer Science\\
  The University of Adelaide\\
  Australia \\
  \texttt{mehdi.neshat@adelaide.edu.au} \\

  }

\begin{document}

\maketitle

\begin{abstract}
 Ocean wave energy is one of the latest renewable energy resources, projected to be commercialized and competitive with other energy technologies in the near future. However, wave energy technologies are not fully developed, so various criteria must be optimized to enter the energy market. To optimize the performance of wave energy converters (WECs) components, three challenges are mostly considered: i)Power take-off systems settings (PTO), ii)Geometry parameters of WECs, iii)WECs' layout. As each of them plays a significant role in harnessing the maximum power output, this paper reviews applied optimization techniques in WECs. Furthermore, due to the importance of fidelity and computational cost in numerical methods, we discuss methods to analyze a WEC together with the basics and developments of WECs interactions. Moreover, the most popular optimization methods applied to optimize WEC parameters are categorized, and their key characteristics are briefly discussed. As a result, in terms of convergence rate, a combination of bio-inspired algorithms and local search can outperform the competition in layout optimization. A review of PTO coefficients and the geometry of WECs have emphasized the indispensability of optimizing PTO coefficients and balancing design parameters with cost issues even though it is costly on multi-modal and large-scale problems.

\end{abstract}
\doublespacing
\sloppy

\section{Introduction}\label{Section1}
Nowadays, due to the necessity of investigating ocean renewable energies, particularly wave energy which has a great potential for energy harnessing, the need for comprehensive studies on this issue has become essential. Interestingly, the number of publications on this subject has skyrocketed over the last two decades. Although the reasons for the increased interest may be environmental and financial, the capacity of extracting energy from optimized wave energy converters is enhanced in comparison with a non-optimized array of WECs. Researchers apply optimization methods to various aspects of the project, from determining the best installation location to minimizing the Levelized cost of energy (LCOE). The optimization methods can firstly be used to identify potential locations for extracting more power from incident waves. Next, after selecting the type and model of WEC, the shape and dimensions of the converter can be optimized. Following that, capturing more power necessitates the use of optimal Power Take-Off parameters and the selection of an effective control strategy for a chosen WEC \cite{82-jusoh2021estimation}, which has a high priority in the European countries. Furthermore, the mooring and foundation of such structures are challenging in which to find the optimal design.
One of the most important optimization problems of exploiting wave energy projects is the configuration of the WECs in an array because the interaction and coupling effects between the converters and water may have a direct relationship with the power output of the array. Researchers have even attempted to optimize the q-factor (a factor that determines whether the interactions are effective or destructive) in order to achieve the optimal configuration and maximize the power output \cite{32-goteman2018arrays}. Obviously, there are other issues in between that were taken into account to benefit from finding the optimal solution.
The problems of extracting renewable energy from waves (which involves using various types of wave energy converters, particularly point absorbers) have majorly been solved over the last three decades. Furthermore, case studies have been published based on potential locations to assess the probability of harnessing energy. Besides that, the importance of finding the optimal solution to the related problems has recently motivated researchers to focus on this issue. In fact, not only has the interest in optimizing studies increased, but also has the interest in solving problems using Computational Fluid Dynamics (CFD) methods\cite{dafnakis2020comparison}.

In recent years, many papers have been published on the optimization problem of wave energy converters, with the majority of modifying and applying metaheuristic optimization algorithms such as Genetic Algorithm (GA) \cite{2-lyu2019optimization} Differential Evolution (DE) \cite{88-gomes2012hydrodynamic},  gradient descent \cite{89-abraham2012optimal}, and Simulated Annealing algorithm (SA) \cite{44-liu2020optimization}. All aspects of the array, such as lifetime cost, Capital Expenditure (CAPEX), LCOE, environmental impact, maintenance, and others, must be considered. However, The maximum power output is the most intriguing parameter for increasing revenue from energy extraction.

The purpose of this paper is to review articles that investigate the optimized configurations of an array of WECs, regardless of whether the optimized fitness function is directly developed for this purpose or not. We attempted to discuss specific aspects of wave energy problems, such as analyzing a single WEC and interaction effects, in a brief manner. Before arguing approaches to analyzing a wave energy converter and discussing interaction resources and parameters, we begin the paper by looking at six types of converters, followed by an argument over worldwide projects and related case studies. Following that, we classified metaheuristic optimization-related articles based on the scientific community's interest. We provide a review for the PTO system, the geometry of a WEC, as well as regular and arbitrary patterns,in three parts.

The structure of this article is outlined as follows. Section \ref{Section1} introduces the classifications for WECs, then provide an overview of recent case studies and projects in different countries.  Section  \ref{Section3} enumerates articles that surveyed the numerical methods for analyzing a wave energy converter and the interactions between the converters in an array. After that, we present the case study papers from recent years. Section  \ref{Section2} introduces the recent technologies and heuristic optimization methods used in the fields, followed by a more in-depth look at the layout configuration, PTO parameter, and geometry of a WEC. Finally, section \ref{Section4} concludes the article.

\subsection{Converters Classification}

Professor Salter \cite{salter1974}, who studied wave power and discussed its potential, pioneered wave energy conversion and the installation of various WECs in 1973. Later, academia and a number of institutions set aside funds to facilitate the research.

\begin{figure}[ht]
\centering
\includegraphics[clip,width=\columnwidth]{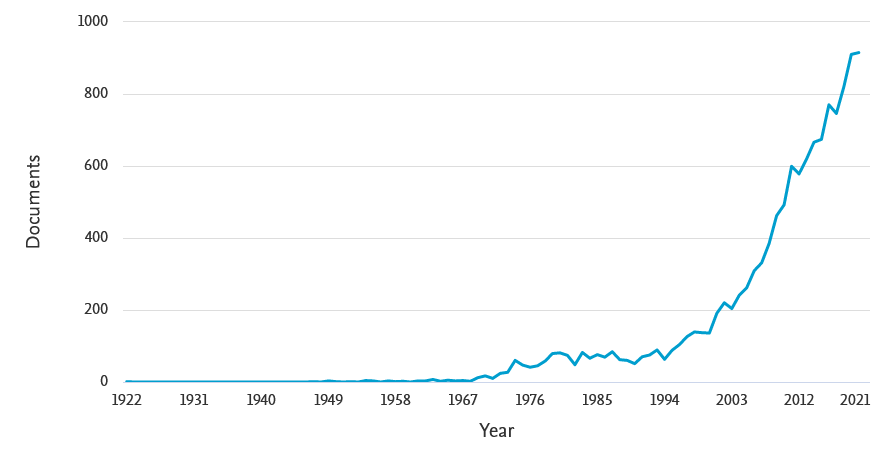}

\caption{The statistical results of published papers in the field of 'ocean wave energy' between 1922 and 2020~\cite{SCOPUS2020} }
\label{fig:statistical_review}
\end{figure}

\begin{figure*}
%[pos=tbp,align=\centering]
\centering
\includegraphics[scale=0.65]{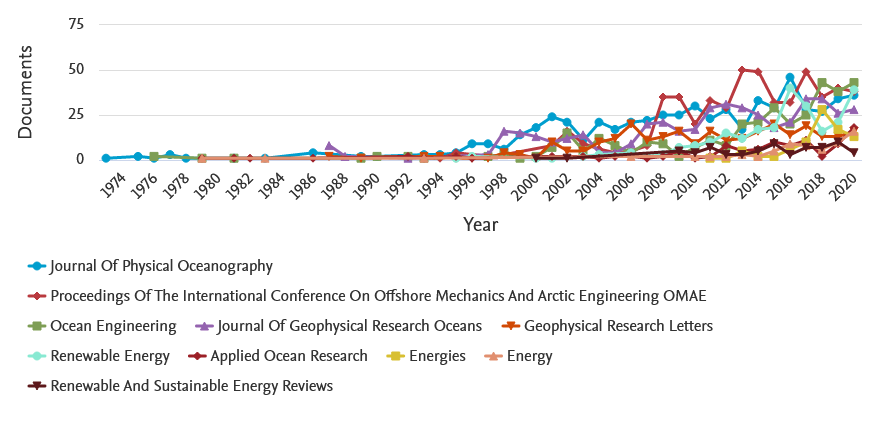}

\caption{The most number of published papers by the top 10 scientific journals in the field of 'ocean wave energy' from 1974 to 2020~\cite{SCOPUS2020} }
\label{fig:statistical_journals}
\end{figure*}

\begin{figure*}
%[pos=tbp,align=\centering]
\centering

 \subfloat[]{
 \includegraphics[scale=.4]{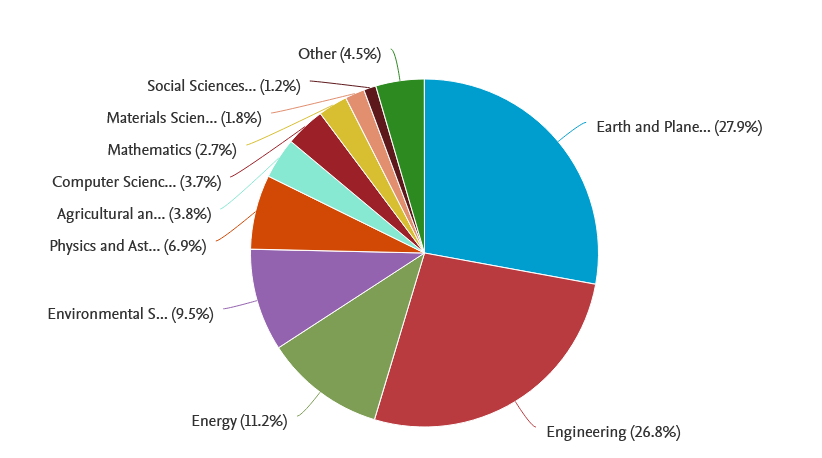}}
 \subfloat[]{
\includegraphics[scale=.4]{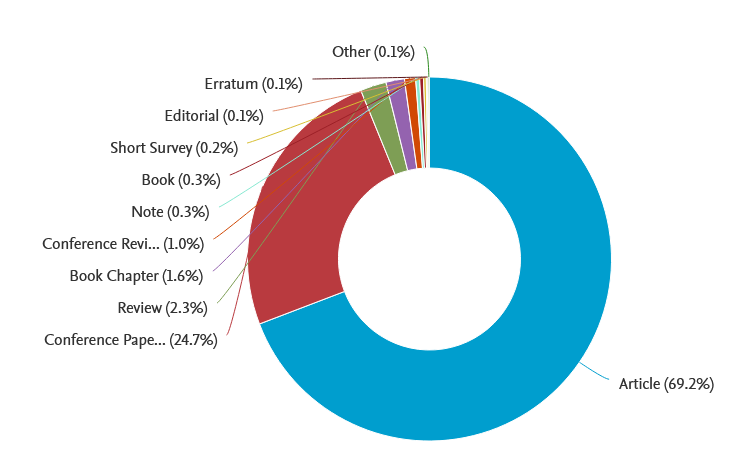}}
\caption{The contribution percentage of (a) subject areas, (b) documents type in the field of 'ocean wave energy'~\cite{SCOPUS2020} based on the published papers. }
\label{fig:statistical_contribution}
\end{figure*}

Figure \ref{fig:statistical_review} shows that in the 1970s, researchers became interested in the field of ocean wave energy. However, since the beginning of the twenty-first century, the number of documents in this field has increased significantly. Figure \ref{fig:statistical_journals} represents the number of publications by the top ten journals in this field, from 1974 to 2020. Between 2000 and 2002, the Journal of Physical Oceanography published approximately 70 papers that made notable contributions to academia. Other journals, such as Renewable Energy and Ocean engineering, have published numerous papers in this field during the last three years.
According to the published papers, various subjects such as earth and planet, engineering, and energy have the most interest among researchers, as shown in Figure \ref{fig:statistical_contribution}. Furthermore, most mentioned subjects are articles and conference papers, while review papers contain only 2.3\% of all published studies.

%-------------------------
%-------------------------
%-------------------------
%-------------------------
%-------------------------
%-------------------------
%%%%%%%%%%%%%%%%%%%%%%%%%%%%%%%%%%%%%%%%%%%%%%%%%%%%%%%%%%%%%%%%%

To this day, there are at least six well-known classifications of converters, based on the installation location, type, principle of PTO system, working principles, floating or submerged, and degree of freedom (DOF) number. The last two items are easy to follow based on their definitions because they are either floating, fully submerged, partially submerged structures, or placed on the seabeds. There are also six degrees of freedom that can be divided into two parts. The first part contains rotational degrees, including roll, pitch, and yaw DOF, and the second one contains translational degrees, including heave, sway, and surge. The WECs can be classified based on the number of DOF they have. It should be note that, a few of WECs are hybrid such as \cite{ren2020experimental} that experiments the concept of combination of tension leg platform wind turbine with a heaving wave energy converter. Here is given a brief description for other helpful classifications.

\begin{enumerate}
% [i)]
    \item Location:
    There are three parts to the installation site for converters. Onshore systems are WECs that are built along the shoreline or attached to constructed structures such as breakwaters. Moreover, Nearshore devices are placed at a depth of 10 to 25 meters, 500 to 2000 meters from the shoreline. Nearshore has Limited bathymetric zones, around a quarter wavelength. Finally, offshore systems, which range in depth from 40 meters to over 100 meters, make selecting an adaptable converter to withstand deep water incoming waves challenging\cite{lopez2013review,babarit2017ocean}.
    \item Type:
    The proportion of WECs' dimension to wavelength, as well as their direction to the incident wave, can be classified into the following categories. Attenuators are large, float-on-the-water devices that are oriented parallel to the wave direction. The next one is terminators, which are installed perpendicular to the predominant wave direction and have a regular size that is greater or equal to the wavelength. The last type is point absorbers, a smaller device than the others and has a shorter incoming wavelength. They can be either floated or submerged converters \cite{8-drew2009review,babarit2011-2019}.
    \item Principle of PTO system:
    The PTO system is one of the most substantial parts of a WEC. Therefore, various methods have been used to produce electricity. Hydraulic motors, turbines with WECs, and electrical drive-based systems are all well-known PTO system methods \cite{28-wang2018review}. \cite{80-ahamed2020advancements} also stated that other PTO systems, such as direct mechanical drive systems, triboelectric nanogenerators, and hybrid systems, have recently been used. The working principles of each PTO system are different from others. For instance, some have an air chamber and convert power from air pressure, especially Wells turbine and impulse turbine. Another type, on the other hand, has hydraulic oil to drive the motor for generating power.
    \item Working principles:
    WECs can be classified into three types based on their functionality. The first is overtopping devices, which absorb energy from waves passing over a ramp that fill a higher level basin or reservoir before releasing the stored water into the sea. The second type is the oscillating water column (OWC). It consists of an air turbine and a chamber with two openings in the bottom and above it. The rise in the water compresses the air in the air turbine to produce energy. The last one is wave activated bodies, which generate energy from the motions of converter caused by the waves \cite{babarit2017ocean}. In this classification, mainly introduced by Falcao \cite{Falcao2010wave}, both PTO system and floating or submerged structures were divided. Moreover, different types of hydro-mechanical conversion are proposed to classify the mentioned three types. This has two Rotating generators with or without energy smoothing, hydro-pneumatic or oleo-pneumatic conversion system, and by electrical generator directly or indirectly from the movement. More information about the hydro-mechanical conversion types is described in section \ref{subsection2.3}.
    
\end{enumerate}

\subsection{Case Studies Assessment}\label{subsection3.3}

\begin{figure*}
%[pos=tbp,align=\centering]
\centering
\captionsetup{justification=centering,margin=1cm}
\subfloat[]{
 \includegraphics[scale=.65]{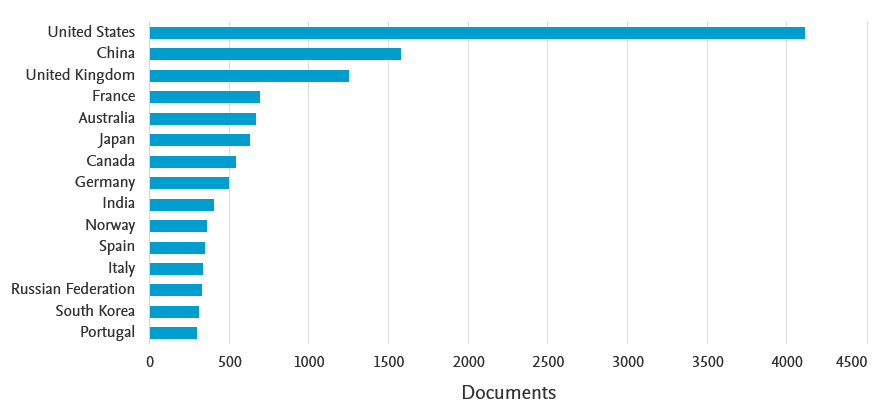}}\\
 \subfloat[]{
\includegraphics[scale=.65]{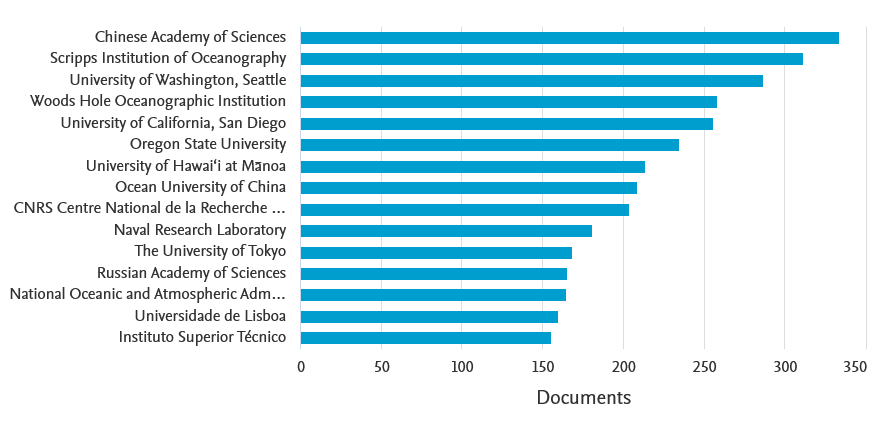}}
\caption{The sorted top (a) countries and (b) institutions with regard to the number of published documents in 'ocean wave energy' topic from~\cite{SCOPUS2020}. }
\label{fig:statistical_countries}
\end{figure*}

During the recent years, countries contributed to publishing documents related to ocean wave energy. However, based on Figure \ref{fig:statistical_countries}, few of these countries' researchers published more than 500 research articles. The United States has worked on this subject and published over 4000 documents, outnumbering China's 1550 papers by far. It should also be noted that institutions can assist academia increase the number of studies. For instance, the Chinese Academia of Sciences and Scripps Institution of Oceanography have collaborated to publish over 600 documents, while the University of Washington Seattle has published approximately 280 papers. Institutions that receive government or private funding can play a vital role in both the industry and academic society.

%-------------------------
%-------------------------

%Other possible research areas are Australia and the United States of America. To give an example,  Ma \cite{40-ma2020research}  apply regular wave data of typical climate conditions of the South China Sea to solve hydrodynamic interactions in the time domain. Bonovas \cite{76-bonovas2020modelling} takes data for an entire year from an online free database called Coastal Data Information Program to survey the design of the PTO system in Monterey Bay in California.(plus china)

%Case studies in the west of Asia are interesting to search the area with high potential of extracting energy from waves. For instance, this possibility is checked by Behzad \cite{hamedbehzad} in the east of the Persian Gulf based on the wave-rose on the installation point from a previous study. Also, Australian coasts were chosen for case studies research projects. Four Coasts of Tasmania, Sydney, Perth, and Adelaide are considered for harnessing the wave energy. Amini \textit{et al.}\cite{amini2020parametric}, and Neshat \textit{et al.} \cite{20-neshat2020hybrid,21-neshat2019new,23-neshat2019adaptive} used these locations' real wave data to find reliable layout of installing WECs. 

This increase in the number of publications led to initiate large projects by countries and companies in the world. Although a few of them in the recent years working productively, in 2019, the ocean energy sector encountered some obstacles, including the sinking of the Wello's Penguin WEC \cite{WelloP} and the termination of the Western Australian government's contract with Carnegie due to the developer's financial difficulties \cite{WAcarnegie}. Nonetheless, the sector is ripe for technological advancement. The devices currently in use have increased survivability by operating throughout the year under extreme marine conditions; However, due to the limited amount of electricity generated by WECs so far, most devices are still in the pre-commercial phase. Improvements must be made in matters such as WEC design optimizations, the validation of PTO reliability, and proposed LCOE targets. In particular, Wave energy extraction in Europe could reach 30 GW by 2050 under the European Strategic Energy Technology Plan cost-cutting scenario \cite{simoes2013jrc}\cite{etipocean}. Also, Scotland, the EU, and the United States have all provided research and development funding to projects aimed at developing low Technology Readiness Levels (TRL) technologies and innovative PTO systems since 2016 \cite{EU2020-report}.

In recent years, researchers have focused on European countries more than other regions such as the United States because not only is the mean annual energy is potentially high in those locations but also prospective vision of the EU facilitate the progress of exploiting renewable energy. As an example, Europe continues are the world leader in wave energy sector, with the most full-scale wave energy devices and 1,250 kW of capacity installed per year since 2010.
Europe now has the opportunity to consolidate its lead and dominate a new global high-value market \cite{etipocean}.

%EU
Since the Horizon 2020 (H2020) Framework program launch in 2014, the European Commission (EC) has funded 47 projects to improve various ocean energy technologies. Projects examples are, Waveboost project working on Corpower C3 device optimization, the LiftWEC project to explore the development of WECs concept based on the exploitation of lift forces generated by wave-induced water velocities, and the Opera project, which employed the Oceantec Marmok device in Spanish waters \cite{Marmok}. The Imagine project aims to create a new Electro-Mechanical Generator intended for wave energy applications, which can reduce CAPEX of current PTO technologies by more than 50\% while increasing average efficiency to more than 70\% and a lifetime to 20 years. The Wave Energy Scotland (WES) PTO program has already provided funding for this technology. The SEA-TITAN project seeks to design, build, test, and validate a direct drive PTO solution that can be used with various WECs. The majority of current wave energy research and development projects are centered on this critical aspect of WECs, the development of reliable PTO \cite{EU2020-report}.

%UK
Supporting such technologies also has been done at a national level. 
The United Kingdom was the first to set the goal of reaching net-zero emissions by 2050. In order to attain this objective, the UK Research and Innovation (UKRI) is currently funding eight projects to work on state-of-the-art wave energy technologies. The main goals of these projects are to enhance the performance and survivability of WECs by developing and testing novel WEC generators, essential device controls and monitoring systems, a highly accurate modeling suite, and investigating the possibility of using new flexible materials for WECs\cite{UKRI}. 
In addition, UK established the WES, which has supported 90 research projects since 2014. Projects were included novel device concepts, new materials, PTO, and control systems of wave energy devices\cite{OES2019-report}.

% Sweden - Portugal 
Corpower, a Swedish company, attracted almost 20 million Euros for a wave energy project in the northern part of Portugal. The first commercial-scale C4 Wave Energy Converter will be deployed off the coast of Agucadoura in 2021 as a part of a four-system WEC array. By noting extracting five times more energy per ton of device from amplified power capture system, they hope to commercialize the technology till 2024 entirely. The point absorber converter uses Bodycote care in order to improve survivability and its resistance against collision\cite{corpower}\cite{corpowernews}. 

%AUSTRALIA
The Australian Renewable Energy Agency (ARENA) has funded Australian companies\cite{OES2019-report}, including Bombora, to investigate the economics of a 60MW wave farm consisting of 40 Bombora WECs at a site near Peniche in Portugal which was completed in 2016\cite{ArenaBombora}, Carnegie to develop the CETO6 device (which was canceled later in 2019) \cite{ArenaCarnegie} and the ongoing Wave Swell project to construct the UniWave200, a 200KW WEC at King island\cite{ArenaWaveSwell}. 

%CHINA
In China, Guangzhou Institute of Energy Conversion (GIEC) researched, developed, and designed Zhoushan, the first 500kW Sharp Eagle WEC, which was officially deployed in Mazhou island's waters in 2020 and is China's largest single-installed WEC\cite{sharpeagle-china}.

%USA
California, Oregon, Washington, Alaska, and Hawaii have promising wave resources for WEC technologies in the United States. However, the Pacific Northwest's abundance of water supplies appears to be preventing widespread adoption in Washington and Oregon. The East Coast area has a less intense wave resource that may suit smaller-scale and distributed applications\cite{BlueEconomy}.
Alaska's wave resource is 890 TWh per year, which is almost 62\% of the total available wave energy resource of the U.S\cite{MarineEnergyUSA}.
Various projects recently succeeded and currently expanding the absorption of wave energy by U.S. companies\cite{EMEC}. One of the companies currently working on wave energy extraction is Oscilla Power. They are developing multi-modal point absorber WEC called Triton for large-scale arrays and Triton-C for remote communities, both of which can capture power from six DOF. The performance is validated via physical testing of different scales performed by the Department of Energy (DOE). This WEC has a low installation cost due to the use of flexible tendons\cite{oscilla}\cite{OES2019-report}.

%%%%%%%%%%%%%%%%%%%%%%%%%%%%%%%%%%%%%%%%%
\section{Recent Challenges and Advances} \label{Section3}
It is evident that as technologies and advances grow, Challenges grow simultaneously. A challenge in surveying wave energy converter projects is selecting a method for analysis of a WEC and solving hydrodynamic interactions. Moreover, having an understanding of acting interactions and applied forces is beneficial.
This section discusses the recent studies about numerical methods to simulate a WEC, followed by challenges over interaction and its parameter.

%%%%%%%%%%%%%%%%%%%%%%%%%%%%%%%%%%%%%%%%%%
\subsection{Individual WEC Analysis}\label{subsection3.1}
In most wave energy converter arrays projects, an isolated WEC must be simulated. There are several ways to simulate the WEC, which are different in terms of simulation time and fidelity. From lowest to highest simulation time, some approaches of modeling are mentioned. The most common Methods among the researchers are potential flow (PF) based models, which itself divided into four models \cite{98-babarit-folley2012review}.
\begin{itemize}
\item  Linearized potential flow in frequency domain
\item  Semi-analytical techniques 
\item Linearized potential flow in time domain 
\item Nonlinear potential flow
\end{itemize}
The first approach is using in most of the studies, and its most popular solvers are WAMIT, NEMOH, ANSYS AQWA, based on the number of published articles. Regardless of limitations in the linearized method, researchers continue to use it especially in small arrays because it returns valuable results in short time \cite{99-babarit2010impact,102-li2020numerical}. Semi-analytical approach has been investigated by many researchers in this field. Some of the most popular solvers used in recent codes are direct matrix method from\cite{direct-matrix-method-kagemoto1986interactions}, multiple body radiation and diffraction\cite{multibody-mavrakos1991hydrodynamic}, multiple scattering either iterative or non-iterative (\cite{ohkusu1974hydrodynamic,multiplescattering,mavrakos1987hydrodynamic}) and WEC-MS \cite{Sergiienko}. The difference between linearized potential flow in frequency domain and time domain is that the latter one has the ability of including transient effects and nonlinear external forces\cite{98-babarit-folley2012review}. Furthermore, nonlinear potential flow considers all non-linearity forces such as viscous drag, flow separation, vortex shedding, and other ones in this model \cite{Review-all-penalba2017mathematical}.

When it comes to the numerical analysis of fluid flows, a branch of fluid dynamics called CFD can solve large and complex problems. To this day most accurate methods of solving Navier-Stokes equations are as follows.
The most accurate method with considering small details is Direct Numerical Simulation (DNS). This method can remove any uncertainty induce by turbulence modeling. It solves all length and time scales of turbulence, so the accuracy of given results is high \cite{DNS-sandberg2015compressible}.
 It should be noticed that from the reviewed papers, we did not find any solver based on this method. However, the chance of using such a method is going to increase with the ever-growing technology and supercomputers \cite{105-windt2018high}.
 
Large Eddy Simulations (LES) is the method that large-scale turbulent structure directly simulates and reserves the model for smaller-scale ones. This may increase the fidelity of the turbulent flow structure. The method tries to neglect the smallest length scale to reduce the computational cost compared to DNS. One of LES's theories is Smoothed particle hydrodynamics (SPH) which Liu recently used as a numerical model in prediction and optimization research \cite{12-liu2020prediction}. Another method of solving Navier-Stokes equations is the hybrid approach RANS/LES. The combination of using two methods for equilibrium or nonequilibrium turbulence may have less cost than the LES method, and it can handle massive separated flows and it is more accurate than the RANS method \cite{LEs-georgiadis2010large}.
 Reynold-Averaged Navier-Stokes (RANS) approache is another one which can be defined as time-averaged equations of motions for fluid flow. Several popular tools are used for analysis of a WEC, such as STAR CCM+, OpenFOAM, SWENSE, and Fluent \cite{coe2016extreme}. As mentined earlier, the accuracy of the RANS method is less than DNS and LES but faster to achieve the results \cite{54-devolder2018cfd}.
 %%%%%%%%%%%%%%%%%%%%%%%%%%%%%%%
 \begin{figure*}
% [pos=tbp,align=\centering]
\centering
%\captionsetup{justification=centering,margin=2cm}

\includegraphics[scale=.85]{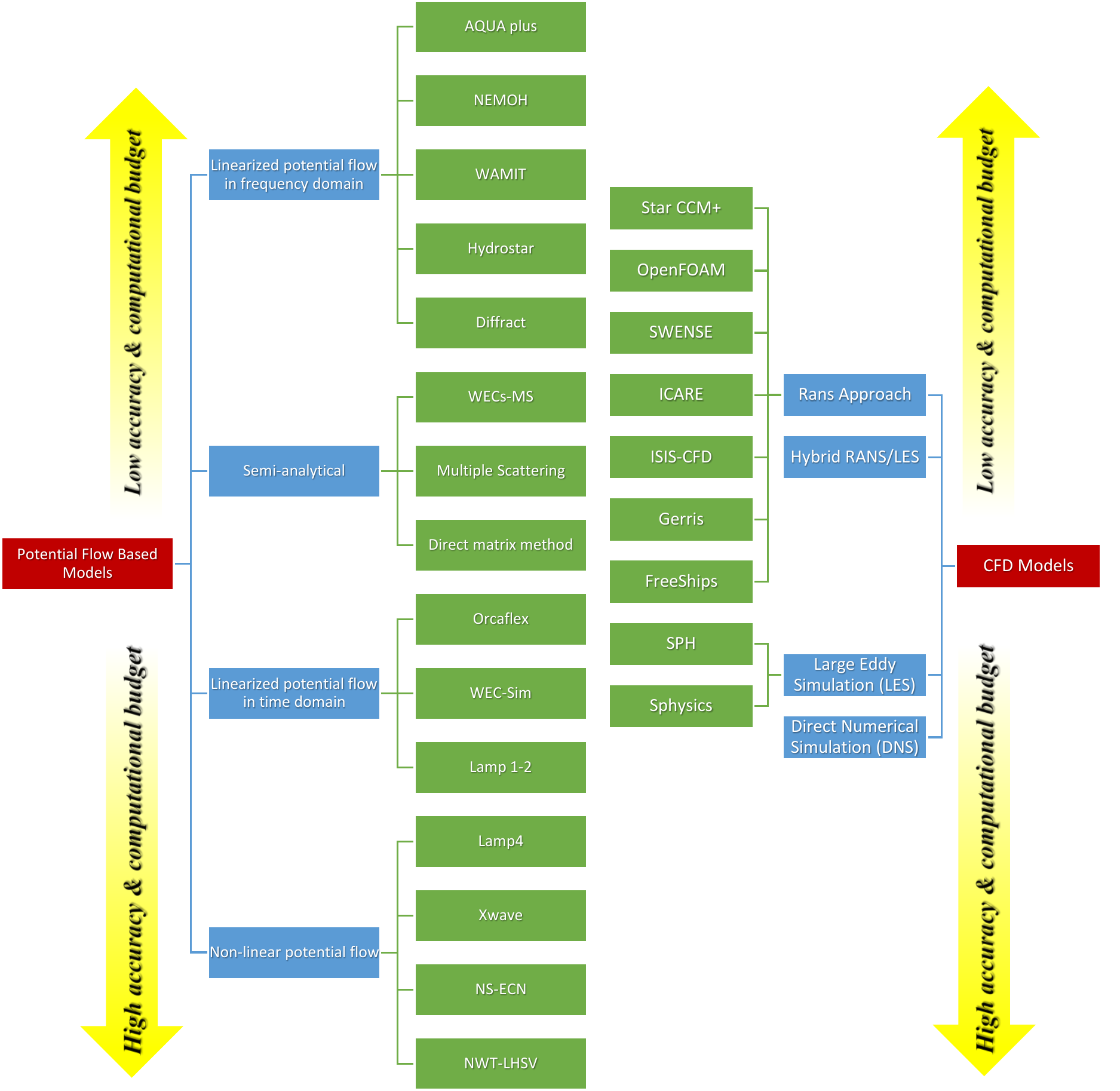}

\caption[centering]{Classification of common-used numerical approaches and related solvers in wave energy converter modeling, and green boxes reveals the numerical solvers. There are more solvers from each approach that is not mentioned above in this figure.}
\label{fig:classification}
\end{figure*}

Figure \ref{fig:classification} uses a tree diagram as an assessment tool to clarify the category of each method. Furthermore, several articles studied and solved the equations with the help of these methods. The Table \ref{table:Models} shows some of the recent articles. It is figured from the table that most of the papers have an inclination to analyze with linearized potential flow models rather than more accurate methods.

 Turning to the factors considered in literature to choose the appropriate method, some of the important factors are briefly discussed. First of all flow must be defined whether as laminar or turbulent. Secondly, in order to simplify the solving equations, flow can be considered as incompressible, irrotational, or inviscid (\cite{104-davidson2020efficient}. As an example, linearized potential flow take all three of them into account. According to \cite{3-goteman2020advances,marchesi2020development}, the most common numerical method for hydrodynamic modeling is Boundary element method (BEM) allowing to numerically solve the motions of WECs having different geometry and shape, with full consideration of wave interactions between bodies \cite{2-lyu2019optimization}. 
 Regarding to BEM solvers, WAMIT is one of the many software for determining the interactions between offshore devices and wave. One of the advantages of this software is having an option to perform high order boundary element method (HOBEM) to improve the computational performance \cite{34-tay2017hydrodynamic}. Moreover, NEMOH is an open source solver, based on BEM codes and has an advantage for the diffraction problem because the code can easily accommodate a user-defined distribution of normal velocities at the centroid of each mesh panel \cite{nemoh-flavia2018numerical}. ANSYS AQWA is a multi-body hydrodynamic program that utilizes three-dimensional radiation/diffraction theory for a global loading and motion simulations \cite{ansysAQWA2013user}.
Needless to say, although the introduced potential flow based models are dominated by researchers in offshore studies, these methods may return unrealistic simulation in case of happening wave resonance because of overlooking viscosity effects \cite{36-goteman2017wave}. And only by using CFD based methods, viscous and turbulent effects can be incorporated \cite{50-bharath2018numerical}. However, the use of CFD tools remained costly for the industry.

\begin{table*}[!htb]
\centering
\caption{ Numerical modeling of wave energy converters array in recent articles. }
\label{table:Models}
 \scalebox{0.9}{
\begin{tabular}{|p{20mm}|l|l|p{34mm}|p{23mm}|l|}

\hlineB{5}

\textbf{Author}   & \textbf{Year} & \textbf{Type of Converter} & \textbf{Numerical Model} & \textbf{Solver} & \textbf{reference} \\ \hlineB{2}
Sharp  &  2017 & Point absorber & linearised PF	&	WAMIT	&	\cite{70-sharp2017array}
           \\ \hline %
           Wu  &  2016 & Point absorber &	Semi-analytical &	Multipole expansion	&	~\cite{78-wu2016fast}
           \\ \hline %
           Lyo  &  2019 & Point absorber & Linearised PF &	NEMOH	&	~\cite{2-lyu2019optimization}
           \\ \hline %
           Hamed Behzad  &  2019 &  OSWEC &	Linearised PF &	ANSYS AQWA	&	~\cite{hamedbehzad}
           \\ \hline %
           Ruiz  &  2017 & Point absorber &	Linearised PF - Semi-analytical  &	Nemoh, Direct Matrix Method	&	~\cite{ruiz2017layout}
           \\ \hline %
            %%%Feng  &  2017 & OWSC &  PF + viscosity & Modified BEM codes	&	~\cite{5-ruiz2017layout}
           %\\ \hline
           Rosenberg  &  2019 & Point absorber &  PF time-domain, RANS & Orcaflex, Star CCM+	&	~\cite{97-rosenberg2019development}
           \\ \hline %
            Parker Field  &  2013 & Oscillating cylinder & RANS approach & Star CCM+	&	~\cite{96-field2013comparison}
           \\ \hline %
           Giassi  &  2018 & Point absorber &	Semi-analytical  & Fast Multiple Scattering  &	~\cite{10-giassi2018layout}
           \\ \hline %
          Finnegan  &  2021 & Point absorber & CFD  &  ANSYS CFX  &	~\cite{finnegan2021development}
           \\ \hline
           Giassi  &  2020 & Point absorber & Semi-analytical, Linearised PF &	Multiple Scattering, WAMIT	&	~\cite{11-giassi2020comparison}
           \\ \hline %
           Bozzi  &  2017 & Point absorber &	Linearised PF time and frequency domain &	ANSYS AQWA, hydrodynamic-electromagnetic	&	~\cite{15-bozzi2017wave}
           \\ \hline %
           Sharp  &  2018 & OWC &	Linearised PF &	WAMIT	&	~\cite{19-sharp2018wave}
           \\ \hline %
           Moarefdoost  &  2017 & Point absorber &	Linearised PF- Semi-analytical &	WAMIT, point absorber approximation	&	~\cite{16-moarefdoost2017layouts}
           \\ \hline %
           Yang  &  2020 & Point absorber &	 PF &	DNV GL SESAM	&	~\cite{18-yang2020wave}
           \\ \hline %
                      Amini  &  2020 & Point absorber &	Semi-analytical &	WEC-MS	&	~\cite{amini2020parametric}
           \\ \hline %
           Neshat  &  2018,2019 & Point absorber & Semi-Analytical & WEC-MS		&	~\cite{22-neshat2018detailed,42-neshat2020optimisation,neshat2020hybrid}
           \\ \hline %
         %   Neshat  &  2018 &  Point absorber &Semi-Analytical & WEC-MS		&	~\cite{neshat2018detailed}
          % \\ \hline %
           %Neshat  &  2020 & Point absorber & Semi-Analytical &	WEC-MS	&	~\cite{neshat2020hybrid}
          % \\ \hline %
          %  Neshat  &  2019 &  Point absorber &Semi-Analytical &	WEC-MS	&	~\cite{neshat2019adaptive}
          % \\ \hline %
          %  Neshat  &  2020 &  Point absorber &Semi-Analytical &	WEC-MS	&	~\cite{42-neshat2020optimization}
         %  \\ \hline %
            Liu  &  2020 &	 OWSC & SPH &	DualSPHysics	&	~\cite{12-liu2020prediction}
            \\ \hline
           Wang  &  2020 & Point absorber &	 PF &	Calculation method	&	~\cite{25-wang2020study}
           \\ \hline %|?
           engstr\"{o}m  &  2020 & Point absorber &	Linearised PF &	WAMIT	&	~\cite{26-engstrom2020energy}
           \\ \hline %
           G\"{o}teman  &  2018 & Point absorber &	Semi-analytical &	WAMIT	&	~\cite{32-goteman2018arrays}
           \\ \hline %
           Tay  &  2017 & OWSC &	Linearised PF, Semi-analytical &	WAMIT, HOBEM	&	~\cite{34-tay2017hydrodynamic}
           \\ \hline %
           G\"{o}teman  &  2017 & Point absorber &	Linearised PF, Semi-analytical &	WAMIT	&	~\cite{36-goteman2017wave}
           \\ \hline %|?
             Ma  &  2020 & Oscillating float &	Linearised PF &	ANSYS AQWA	&	~\cite{40-ma2020research}
           \\ \hline 
           Balitsky  &  2014 & Point absorber &	Linearised PF &	WAMIT	&	~\cite{49-balitsky2014control}
           \\ \hline %|?
            Bharat  &  2018 & Point absorber &	RANS, Linearised PF &	Star CCM+ 	&	~\cite{50-bharath2018numerical}
            \\ \hline %
           Devolder  &  2018 & Point absorber &	RANS, Linearised PF &	OpenFoam, WAMIT 	&	~\cite{54-devolder2018cfd}
           \\ \hline %
             Faraggiana  &  2019 & Point absorber & PF &	Nemoh, WEC-Sim	&	~\cite{56-faraggiana2019design}
           \\ \hline %
            
                       Monroy  &  2011 & -- &	RANSE   &	SWENSE	 &	~\cite{monroy2010rans}
           \\ \hline %

                    \\ \hline

 \hlineB{5}

\end{tabular}
}
\end{table*}

%%%%%%%%%%%%%%%%%%%%%%%%%%%%%%%%%

\subsection{Array WECs interaction}\label{subsection3.2}

\begin{figure}[ht]
\centering
\includegraphics[clip, width= 13cm]{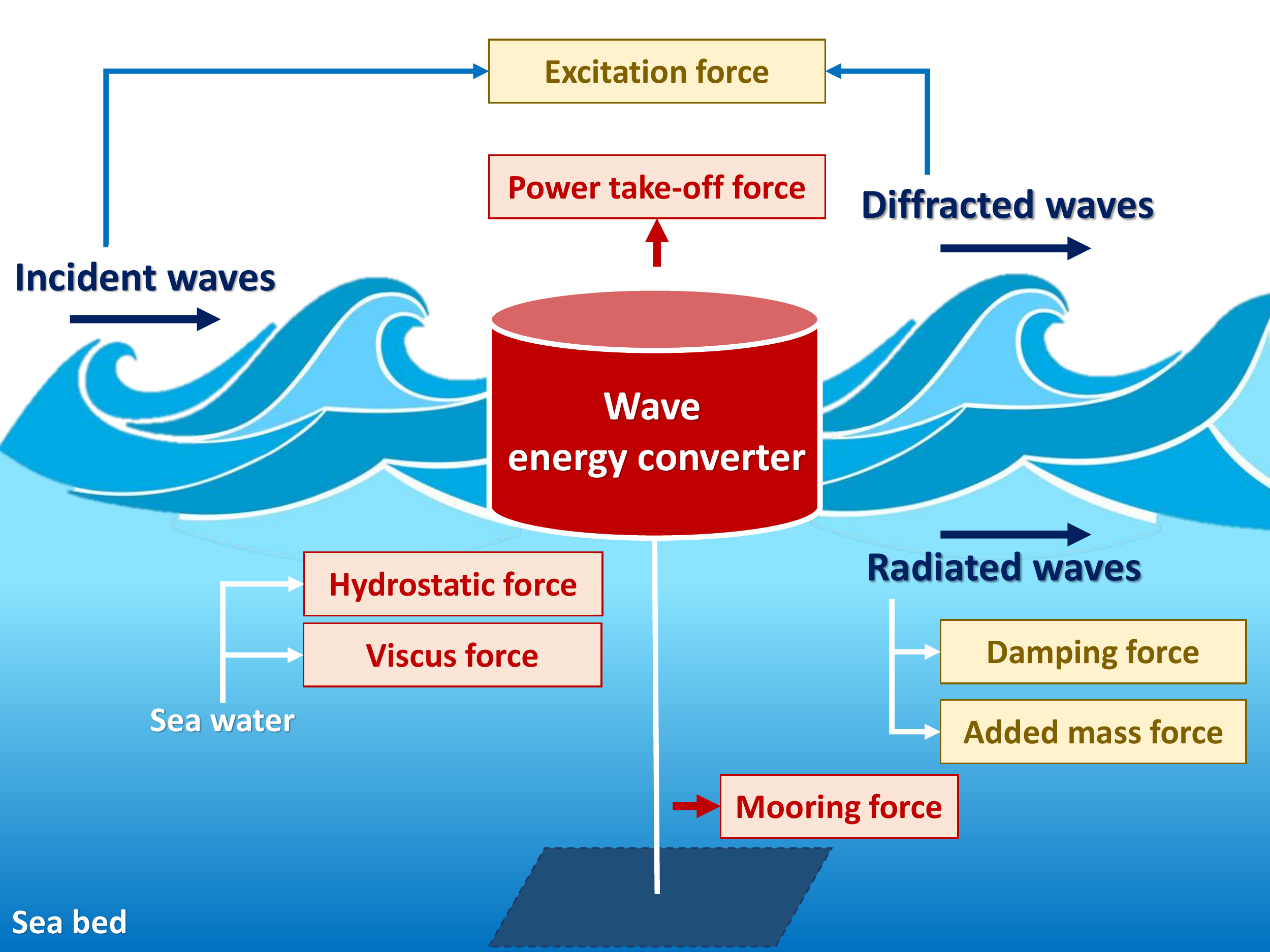}

\caption{An illustrative look on the acting forces on a wave energy converter}
\label{fig:forces}
\end{figure}

In general, waves generated from two sources, winds and swells. The wind or sea waves results from local wind condition, however, swell comes from the storms and winds from away fields. Although the definitions of them are quite similar, there are differences such as low steepness in swell waves to avoid breaking in deep seas, and shorter wave frequency of wind waves compared to the swell waves. The combinations of these two waves form the incoming waves and  affect the converter's motion considerably. The incoming waves or incident waves impact the first device, then it causes a displacement in the device, subsequently the motion resulted from the device creates the radiated waves. Now there are two waves affecting the second device which is producing its own radiated power, therefore this chain of waves occur with the possibility of affecting on the former device. Noting that the PTO parameters are related to the radiation properties of each device, and by optimizing these parameters, more energy will be extracted. Moreover, after the wave hits a converter, wave diffracts and due to this phenomenon a spectrum is made around the converter representing the amplitude of the wave. This may whether increase or reduce the wave height. Such wave phenomena interact and make a difference on the absorbed power and arrangement of the converters. For example, diffracted and incident waves creates the excitation force, and similarly radiated waves have relations to added mass, damping force.
As we discussed in \ref{subsection3.1}, selecting an approach to simulate and solve hydrodynamic coefficients, result in calculating applied forces on each WEC. When in potential flow solvers for instance, hydrodynamic coefficients such as added mass, damping coefficients are solved, all forces acting on a WEC device can be calculated. According to the Newton’s second law, which is mass of the body multiplied by acceleration is equal to the acting forces, various forces play a part in solving the equation of motion; Forces such as the excitation force, radiation force, PTO force, damping force, mooring force, hydrostatic restoring force, and viscous force, which can be seen in Figure \ref{fig:forces}. Based on the discussion in \cite{penalba2017mathematical} forces such as Froude-Krylov force, which is introduced by the unsteady pressure field generated by undisturbed waves, and additional force such as drift or wind or other body-water interaction, are also acting on the body. To be more precise, by using the nonlinear Froude-Krylov force together with a quadratic viscous model would result in more accurate solutions of hydrodynamic interactions problem in heaving point absorber \cite{penalba2020systematic}.
Furthermore, other nonlinear phenomena are influencing the process of harnessing energy. For instance, sloshing is relative to the enclosed water, or slamming is coming from the impact of device on the surface of water. Therefore, based on the type of converter, relative forces should compute and consider to raise the accuracy. It is clear that some of the discussed forces can be neglected whether the converter is floating or submerged, point absorber, attenuator or terminator. Then, after calculating the acting forces, the outcome of the interaction on each buoy should be determined. 

%%%%%%%%%%%%%%%%%%%%%%%%%%%%%%%%%%%%%%%%%%%%%%%%%%
%%%%%%%%%%%%%%%%%%%%%%%%%%%%%%%%%%%%%%%%%%%%%%%%%%
\subsubsection{Interaction Parameter}

 When it comes to install more than one buoy into the marine selected area, the interaction between them has to be considered.  The influence of interaction may be seen directly in reduction or increase in the sum of each individual converter power in comparison with the array power in some cases. The ratio between the total power absorbed in an array to the power would be captured by a buoy, is the definition firstly introduced by Budal in 1976 with a parameter called q-factor\cite{budal1977theory}.

\begin{equation}
    q= \frac{P_{array}}{N* P_{individual}}
    \label{budal}
\end{equation}
 where in equation \ref{budal}, $N$ represents the number of converters and $P_{individual}$  stands for the isolated WEC power and $P_{array}$ power from the array.
 
 During the last years many research projects studied  relative equations and optimization of this parameter. There are some publications  related to the calculation of interaction such as the study of Borgarino \cite{51-borgarino2012impact} that used the $q_{mod}$ instead of the traditional equation. Moreover, Fitzgerald \cite{fitzgerald2007preliminary}  introduced a new consistency condition for q-factor  pointed out that there is a symmetry in this parameter  in all directions of the incident waves. This can be described that if in a direction, constructive interaction is observed there would be another direction that cause a destructive one around the converter. Furthermore, another presentation by Wolgamot \cite{wolgamot2012interaction} investigated the wave energy flux together with the wave number, have correlation to the sum of power in all directions of an array. It should be noted that this is proved by excluding the assumption of shape and size of the body.
 
 \begin{equation}
    q_{mod}( \omega) = \frac{P_i(\omega)- P_0(\omega)}{max_{\omega}(P_0(\omega))}
    \label{q-mod-borgarino}
\end{equation}
where $P_0$ and $P_i$ is the power output of an isolated WEC, and the power captured from i-th body in the array, respectively in equation \ref{q-mod-borgarino}.  Following that, other researchers tried to represent a new related definition or an analytical approach for the purpose of determining the effectiveness of interactions.  Sun  \cite{47-sun2016linear} gives an extracted relation based on the former q-factor. They considered different distances and wave periods for observing the potential of interaction effects.
\begin{equation}
\label{equation:q-sun}
    q= \frac{ P_{c,array} - N_{array} P_{c,isolated}}  {N_{array} max\left[  P_{c,isolated} \right]} 
    +1
\end{equation}

  where in equation \ref{equation:q-sun} maximum of $P_{c,isolated}$ is the highest measure of the absorbed power by an isolated WEC.
Likewise of equation \ref{budal}, same definitions of q-factor works here and if q-factor is larger than one, constructive interaction happen.  
A recent publications \cite{25-wang2020study} represents the new evaluation index of array energy gain coefficient. This method can calculate the efficiency of layout by studying the hydrodynamic characteristics.

 Researchers also investigated the optimal system for maximising the q-factor to benefit from the constructive interaction and its relative merits. It is believed that by having constructive interaction, array will perform efficiently with higher energy absorption. Therefore, many researchers tried to discuss the results of the WECs in an array by reaching the highest q-factor. For instance, Moarefdoost \textit{et al.} \cite{16-moarefdoost2017layouts} used heuristic techniques to  maximise the value of analytical q-factor with regards to finding the distance and direction in polar coordinate ($d$ and $\alpha$) . Another case \cite{2-lyu2019optimization} tried to maximise the q-factor together with 3 design variables, and assume some constraints in the optimization process solving by GA. Also in the mentioned paper, verification of the conclusion from case study of Moarefdoost is studied, which claims that the q-factor will not change noticeably if multiplication of wavenumber to dimension of the WEC ($kR$) remains the same. Finally According to table \ref{table:layouts}, it can be witnessed that some of the papers  optimized the q-factor and the results of them show direct relation to layout, energy absorption, and even the cost of project.
 
Two of the most important results of having constructive interactions are as follows. For instance, it is considered in \cite{32-goteman2018arrays} that the power output is influenced by the hydrodynamic interactions between devices, and also assessed the wave direction in this study. The understanding of optimal layouts of WECs is improved by using optimization methods aiming to maximise the interaction factor \cite{19-sharp2018wave}.

\section{Review of the Recent optimization Studies and Developments}\label{Section2}

In this article, we attempted to review optimization algorithms, with a focus on the development of layout-related problems, PTO systems, and geometry studies of WECs in recent years. 

%%%%%%%%%%%%%%%%%%%%%%%%%%%%%%%%%%%%%%%%%%%%%%%%%%%%%%%%%%%%%%%%
\subsection{Applications of optimization techniques}\label{subsection2.1}

Meta-heuristic optimization algorithms are extensively applied to find the best multiple optimal solutions (in a single run) out of all possible solutions in solving multi-objective optimization problems. They evaluate potential solutions and implement a series of operations to them in order to generate different offspring, and hope to converge the better solutions. Over the last decades, many meta-heuristic algorithms have been developed to solve a wide range of real world engineering optimisation problems\cite{sorensen2013metaheuristics}. We have listed meta-heuristics into three classes based on previous studies, including numerical methods, bio-inspired algorithms, and local search methods.
Many algorithms combine ideas from different classes, which are called hybrid methods \cite{sorensen2013metaheuristics}.
Researchers have also conducted comparative studies to evaluate optimization methods' performance. Such studies can be highly beneficial in assisting end-users to select the best optimization method for their problems\cite{beiranvand2017best}.

Given the plethora of optimization methods and their implementations in related papers, various optimization methods have been proposed to improve the energy extraction capability of WECs. The given problem, the priority of main decision parameters,  and the design of the objective function are all crucial in determining the best optimization practices.

%To obtain the optimal layout of WEC arrays, Ruiz {et al.} introduced a four-parameter layout description and compared three optimization algorithms of covariance matrix adaptation evolution strategy (CMA-ES), a GA, and the glowworm swarm optimization (GSO) algorithm. Their research revealed that, while the GA and GSO performed slightly better, the CMA was noticeably less computationally demanding \cite{ruiz2017layout}.

\subsubsection{Numerical Optimization Methods}\label{subsubsection2.1.1}
%such as 1) Nelder-Mead 2) Gradient descent 3) Conjugate gradient 4)COBYLA
Although converging to the best solution directly in the majority of real engineering problems is challenging, it is relatively simple to set up a loss function that measures the quality of a solution and then minimize the function's parameters to find the optimum solution that should be feasible~\cite{nocedal2006numerical}.
For instance, one of these algorithms is gradient descent which minimizes the cost function as far as possible.

Some of the research projects investigates numerical methods which are discussed further. Noad and Porter \cite{68-noad2015optimisation} adopted a multi-dimensional numerical optimization procedure for single flap-type OWSC to evaluate the solution method's accuracy and then to set up optimal device parameters for array optimization. To obtain the optimal layout of WEC arrays, Ruiz {et al.} introduced a four-parameter layout description and compared three optimization algorithms of covariance matrix adaptation evolution strategy (CMA-ES), a GA, and the glowworm swarm optimization (GSO) algorithm. Their research revealed that, while the GA and GSO performed slightly better, the CMA was noticeably less computationally demanding \cite{ruiz2017layout}. Moreover, Raju \cite{79-rajuthesis} used the derivative-free continuous optimization method, CMA-ES, to optimize the converters' placement in terms of minimizing their negative interactions and the Nelder-Mead (NM) search algorithm to find the optimal PTO parameters for each WEC.

In order to maximize the total absorbed power output of the WECs, Abraham \textit{et al.}\cite{89-abraham2012optimal} formulated an optimal control method for a PTO mechanism that includes linear dampers and active control elements. The used solution was a bang-bang type. Based on the results, the gradient projection method was inexpensive and more feasible compared to a general nonlinear program (NLP) solver.

A shallow artificial neural network (ANN) is presented in Thomas \textit{et al.}'s study \cite{71-thomas2018experimental} to determine optimal latching times in irregular wave conditions. Based on the simulation, for some sea states, the learnable WEC absorbs more than 30\% power compared to the best constant latching time in the test wave and absorbs a double amount of the power of the WEC without latching. Furthermore, the latching–declutching optimization is a sophisticated PTO control technique in order to improve the performance of the whole system. Since the cost function in this optimization problem is discontinuous, Feng and Kerrigan decided to apply a novel derivative-free coordinate-search algorithm, using a formulation based on the past wave information and prediction of the future wave. The algorithm was compared with a derivative-free global meta-heuristic method, the simulated annealing algorithm (SA), to prove the efficiency \cite{92-feng2015latchingdeclutching}.

\subsubsection{Bio-inspired optimization algorithms}\label{subsubsection2.1.2}

Bio-inspired algorithms solve complex problems by employing simple nature-inspired methods. This approach mimics nature's strategy, as many biological processes can be viewed as constrained optimization processes. They employ a large number of arbitrary decisions, so they are classified as a subset of randomized algorithms. This approach's popularity stems from the fact that it is helpful for solving a wide range of problems and can be applied to complex problems in all key areas of computer science \cite{binitha2012bio}. The bio-inspired algorithms have been classified into evolutionary-based algorithms (EA), swarm-based intelligence algorithms, ecology-based algorithms, and multi-objective algorithms \cite{fan2020review}. Here we discuss two of these classes: the most predominant classes amongst others—Evolutionary Algorithms and Swarm-based Algorithms, inspired by the natural evolution and collective behavior in animals, respectively \cite{binitha2012bio}.

\begin{itemize} 
\item Evolutionary-based algorithms:
Evolutionary algorithms searches for good answers in the search space. The typical procedure of such algorithms are threefold. At the beginning, they initialized the samples, then objective function is calculated together with the selecting operation between the former and new ones. and use termination criteria for reducing the computation, after transforming selected samples to additional ones.

%GA - genetic programming (GP) - Evolution strategies ((1+1)-EA, ($\mu$ + $\lambda$)-EA, ...) - differential algorithms (DE) - Paddy Field algorithm 

To overview of using evolutionary algorithms in the subject of wave energy converter, we start by an early optimization study \cite{91-child2010optimal}. Parabolic Intersection (PI) and GA methods are used to find optimal array configurations in early WEC optimization studies. Despite the fact that GA outperformed the PI in terms of performance, the PI was much more computationally efficient.
A similar research is done by Moarefdoost \textit{et al.} \cite{16-moarefdoost2017layouts} that proposed a heuristic optimization algorithm to find a layout with maximum q-factor. The results of this algorithm outperformed the modified GA.
Related research projects by GA can be seen in the below arguing papers. Lyu\textit{ et al.} \cite{2-lyu2019optimization} used GA to optimize both dimensions of individual WECs and the array layout of the WECs' cylindrical buoys.
 A hidden genes GA (HGGA) is suggested by Abdelkhalik and Darani \cite{abdelkhalik2018optimization}  for the nonlinear control of the WECs and to optimize system nonlinearities caused by shape, large buoy motions, and the PTO. Results showed that the shape-based approach used in this paper has a reasonable rate of reaching convergency, and it can include all the wave data for optimizing the process.
Giassi and G\"{o}teman introduced a tool according to GA, for optimizing some of the principal parameters such as the value of draft and radius of a single point-absorbing WEC  \cite{59-giassi2017parameter}. They extended the tool to optimize arrays of 4 to 14 similar point-absorber WECs \cite{giassi2018layout}.
Following, Giassi \textit{et al.} \cite{60-giassi2017multi} developed and used the method for arrays of non-homogeneous WECs to obtain optimal wave power parks.
Sharp and DuPont \cite{19-sharp2018wave} presented a GA approach with an additional objective function in this study, the interaction factor. This paper studied the spacing effects and minimized the destructive interactions in order to maximize the power. A beneficial use of machine learning approaches is introduced by Liu. In \cite{12-liu2020prediction}, the study of OWSC started with 9 random design parameters which were simulated by DualSPHysics software. Then, after the capture factors were obtained from training radial basis function neural network (RBFNN), these parameters optimized by GA.
 
%41-parinello?

Although interests in utilizing the GA is high, other optimisation methods have been studied and absorbed a considerable number of researchers in the last decade. For instance, Wu \textit{et al.} \cite{78-wu2016fast} employed two EAs of (1+1)-EA and CMA-ES for optimizing a three-tether submerged buoy array. The combination of the aforementioned algorithms proved to be worthwhile. That is, the (1+1)-EA was used for converging on a solution, and CMA-ES was utilized in fine-tuning that solution. However, the applied wave model in the Wu \textit{et al.} study was not advance and near to the realistic sea wave.

Fang \cite{fang2018optimization} introduced the concept of an adaptive mutation operator to modify the DE algorithm for layout optimization of the three different layouts of point-absorber WECs under a regular wave. The modified DE algorithm turned out to be more functional (both in calculation convergence speed and achieving a better optimal result) than the traditional DE algorithm with a constant mutation operator; however, the adaptive DE was not compared with the popular and modern adaptive, and self-adaptive DE such as SaDE~\cite{qin2005self}, JADE~\cite{zhang2009jade}, EPSDE~\cite{mallipeddi2011differential} and so on . Bonovas et al. \cite{76-bonovas2020modelling} optimized construction parameters by EAs (integrated into EASY software) with the total investment cost and the reservoir flow rate as the objective functions.

An efficient analytical wake model with high accuracy and an energy output model for the OWSCs is introduced by Liu \textit{et al.} \cite{7-liu2021proposal}. Using the proposed models and DE algorithm for OWSC layouts optimization showed that the staggered layouts were economical and appropriate to maximize the total energy power outputs. However, \textit{et al.} \cite{7-liu2021proposal} do not discuss the application of the advanced mutation and crossover operators in the studied DE.

In one of the initial studies, Michael J. D. Powell invented constrained optimization by the linear approximation (COBYLA), a numerical optimization method for constrained problems where the derivative of the objective function is unknown\cite{powell1994cobyla}. 
 Gomes \textit{et al.} \cite{88-gomes2012hydrodynamic} optimized the dimensions of the floater and tube of an OWC to achieve the maximum wave energy extraction. The algorithms used in this geometry optimization were DE and COBYLA. Based on their study, the relatively large variations in the turbine's damping coefficient had a small influence on the annual average power.
Silva \textit{et al.} \cite{eSilva2016hydrodynamic} proposed two optimization algorithms that do not require the function gradient, GA and COBYLA, to solve a typical multi-dimensional, single-objective problem. This hydrodynamic optimization method consisted of the main core and also an internal method integrated inside the main one. The principal framework optimized the floater geometry using GA and COBYLA. Sequentially COBYLA algorithm was used in the internal optimization problem to optimize the turbine characteristics and mass distribution.

\item Swarm-based intelligence algorithms:
Swarm intelligence (SI) algorithms are valuable because of their flexibility in various problems and ability of strong global search. One of the well-known algorithms is particle swarm optimization (PSO). This is a population-based algorithm which are following the direction of particles movement to find local and global optimum solution. Some of the research projects involving SI mention briefly as follows.

A GA, a PSO, and Hybrid Genetic PSO (HG-PSO), were used to investigate the main WEC parameters' values optimization, which the latter one turned out to be the most proper algorithm\cite{capillo2018energy}.
Faraggiana\textit{ et al.}\cite{56-faraggiana2019design} compared the evolution of the minimum LCOE of WaveSub WECs using a GA and a PSO. The results showed that both algorithms operated almost equivalent.
See \textit{et al.} \cite{81-see2012ant} introduced the application of bio-inspired Ant Colony Optimization (ACO) metaheuristic to optimize the electric PTO for point absorber WECs based on the instantaneous changes in the wave pattern. This led to a notable decrease in the computation time, which is valuable as it allows the WEC to achieve resonance in heave (oscillation) with ocean waves.

\end{itemize}

\subsubsection{Local Search Methods}\label{subsubsection2.1.3}
%Such as 1) Hill-climbing search  2) Simulated Annealing 3) Tabu search  4) Local Beam Search 5)Stochastic Local Search 5)NLPQL
Local search metaheuristics find good solutions by making small changes to a single solution iteratively. Hence adjacent solutions are relatively close to each other. By applying a single move to a given answer, the set of solutions can be achieved. A solution from the neighborhood replaces the current solution in each iteration \cite{sorensen2013metaheuristics}.

A stochastic optimization approach was applied by Tedeschi \textit{et al.} \cite{87-tedeschi2013stochastic} to optimize energy storage system sizing to reduce the final cost of energy.
The simulation results of Jusoh \textit{et al.} \cite{82-jusoh2021estimation} reveal that the average electrical power produced from the hydraulic power take-off (HPTO) units optimized by the non-evolutionary Non-Linear Programming by Quadratic Lagrangian (NLPQL) and GA raised to 96\% and 97\%, respectively, in regular wave conditions. Since the NLPQL is a local search method, it is much faster but less reliable than GA. In WECs with data-driven linear generator configuration type, the Hallbach linear generator is used in the secondary structure to minimize the energy loss. To increase this generator's efficiency, the SA was used by Liu \textit{et al.} \cite{44-liu2020optimization}.

Neshat \textit{et al.} \cite{22-neshat2018detailed} started the research of comparing different metaheuristic algorithms. It is shown that using a hybrid method containing stochastic local search and  NM Simplex returns better answers. Moreover, analyzing the results from a random search, partial evaluation, CMA-ES with different settings, DE, iterative local search, local sampling together with NM showed that there is only a 20 percent difference in the mean output of the mentioned methods. After that, the study continues in \cite{neshat2019new} where the buoy positions are surveyed through a new hybrid method that combines local search method with a numerical optimization method. Methods in two groups of population-based and single solution optimizers are used for comparing the results of this hybrid method which utilizes a knowledge-base surrogate power model. The most effective methods to gain more power output are smart local search with or without NM and improved smart local search, combined with NM, sequential dynamic programming, interior point search, and active set search. The last one performs the best among all of the optimization methods. The author mentions the absence of backtracking optimization, and this concern is solved in the following two papers. In \cite{23-neshat2019adaptive}, an adaptive neuro-surrogate optimization (ANSO) method has represented that benefit from a trained Recurrent Neural Network (RNN). Between the tested algorithms, ANSO-S4-B performs better in three of four tested locations.
Furthermore, this algorithm is 3.6\% better in terms of layout performance comparing to local search plus Nelder-Mead (LS-NM). Neshat \textit{et al.}\cite{20-neshat2020hybrid} introduced a hybrid Cooperative Co-evolution algorithm (HCCA) to optimized power output by considering the positioning and PTO parameters for each WEC. This algorithm includes three methods which are SLPSO, SaNSDE, and AGWO, with the same population to solve the problem cooperatively, as well as benefiting from backtracking optimization algorithm. Noting that HCCA responds better than other algorithms, however, in 4-buoy layout, it is not the only best answer.

In conclusion, it is clear that researchers are moving forward to use optimization algorithms in order to find the optimal solution. In spite of that, Finding the best frameworks is difficult. We believe ANSO, and HCCA give viable answers in optimization studies, especially the study of layout and PTO. Furthermore, excessive studies on different GA algorithm, despite the good results, prevent the academic society to explore other algorithms. On the other hand, swarm intelligence algorithms were not surveyed thoroughly. Finally, we suggest to consider multi-objective functions with the recent mentioned algorithms.

%%%%%%%%%%%%%%%%%%%%%%%%%%%%%%%%%%%%%%%%
\subsection{Layout Configuration} \label{subsection2.2}

 Many research projects studied the possibility of increasing energy absorption of the array by considering different layouts. Such layouts may decrease 30 percent of the array or even cause a rise around 5 percent \cite{CChild2011development}. Throughout the last decade, many papers surveyed both simple regular patterns and arbitrary ones. However, optimized patterns may be a reliable choice in order to have constructive effects on extracting power \cite{20-neshat2020hybrid,16-moarefdoost2017layouts,78-wu2016fast}.
 
Several factors have effects on the positioning of buoys in the array. They may have a positive or negative impact based on the overall goal of a project. To be more specific, a shorter distance between WECs mostly increases the destructive effects while decreasing the cabling costs. Six of the most important factors investigated in the layout papers are introduced briefly. First, the number of WECs directly determines the layout configuration. Using more than four converters allows investigating polygons and circular layouts. Second, the separation between each converter can alter the array's interaction effects. Suppose the separation increases to a large number. In that case, the array's interaction effects could be negligible, and the power absorbed by an array would be equal to the power of isolated powers\cite{de2014factors}. Third, the sea state is critical when it comes to the power absorption parameter. When the system's natural period is close to the wave period, wave interaction effects work productively, and a considerable amount of power is absorbed, According to \cite{99-babarit2010impact}. Furthermore, in the case of regular waves, both wave direction and wave frequency directly affect the excitation force, and the excitation force acting on the device has a significant impact on the interaction factor\cite{34-tay2017hydrodynamic}. Fourth, wave direction can be important in determining layout, and it can be investigated by rotating the array pattern or theoretically using different definitions throughout the study. The majority of papers consider unidirectional direction, whereas \cite{32-goteman2018arrays} provides relationships and descriptions for multi-directional waves. Fifth, the extracted power is affected by the size of each WEC, as discussed in section \ref{subsection2.4}. Finally, considering the interaction between waves and devices is essential when positioning the buoys to avoid destructive effects. This will be described in section\ref{subsection3.2} \cite{64-goteman2014methods,70-sharp2017array}.
 
On two levels, there are various layouts evaluated with the aforementioned aspects in array configurations. First, publications cover regular patterns for arrays such as linear, circular, arrow shape, rectangular or square shape, staggered, random, polygon, hexagonal, and other configurations. Second, the urge to extract as much energy as possible from any pattern condition motivates researchers to use optimization algorithms to determine the best layout for converters.

G\"{o}teman \textit{et al.} \cite{64-goteman2014methods}  investigated the park layout as the number of converters increased from 4 to 64. According to this study, increasing the number of converters in the park reduces the power of each WEC. Furthermore, a survey of two wave periods with a 1-second difference revealed that the shorter period captures more power output. It should be noted that the configurations were mostly rectangular, with a semi-circular layout being used to compare the results. In the study of Bosma \cite{39-bosma2020array}, who outlined numerical and physical array testing, optimized and non-optimized layouts with 5 OWCs were considered. The results showed that when the layout is optimized, the average power in regular and irregular waves increased by 12\% and 7\%, respectively. In the study of Amini \textit{et al.} \cite{amini2020parametric} four regular layouts (namely the linear triangular square and pentagon) in four locations in Australia were assessed under different separation and dominant wave directions. The most harnessed energy was in a linear configuration with a separation distance of 165 meters. Baltisky \cite{balitsky2017assessing} concluded that the separation between WECs should be increased to reduce interactions for a constant wave period. To test the application of a generic coupling methodology, Verbuggle \cite{73-verbrugghe2017comparison} tested ten programs consisting of one to five WECs in various arrangements, wave depths, and bathymetries. Garcia \textit{et al.} \cite{garcia2015control} investigated control-influenced array layout,  which allowed wave farms to absorb more power with 2-3-4 WECs in the linear, triangle, and square configurations. It was demonstrated that array performance could be improved by up to 40\%. Raghukumar \textit{et al.} \cite{raghukumar2019wave} investigated the rectangular array layout and separation between WECs to absorb maximum power while minimizing environmental effects. It is concluded that when WECs are placed close to each other, shadowing effects occur, and absorbed power decreases as a result. Giassi \cite{6-giassi2020economical} examined park layout and concluded that for parks with fewer than 20 converters, converters should be placed perpendicular to the wave direction to reduce cable length. Deandres \textit{et al.} \cite{de2014factors} takes into account interactions factor in linear, triangular, and square layouts with WECs ranging from 2 to 4. In this case, it was demonstrated that linear geometry has destructive interaction efficiency, and the other two have similar efficiencies in terms of sea state conditions. Liu \cite{7-liu2021proposal} investigated various staggered layouts of OWSC with varying dimensions and optimized them using the DE algorithm. The mean capture factor of an array was discussed in \cite{68-noad2015optimisation}. The highest efficiency for 3 WECs layouts and 5 WECs layouts was increased by 5 and 7 percent, respectively, compared to a single device's performance. In this study, bowl-like or chevron patterns had the greatest positive increase in power absorption. Ruiz \textit{et al.} \cite{30-lopez2018methodology} studied the performance of different configurations during the life-cycle of WECs, intending to maximize average extracted energy during that period of time while also minimizing interaction effects between WECs. They found that the arrow layout is the most efficient, absorbing up to 20\% more energy than the other geometries. Moarefdoost \textit{et al.} \cite{16-moarefdoost2017layouts} surveyed symmetric and asymmetric layouts and used a heuristic optimization to deduce symmetric layouts answers better except for layouts with four WECs. Using GA in \cite{59-giassi2017parameter} resulted in a spatial layout with no negative interactions. Furthermore, they believe that this may be unimportant in terms of the number of WECs. G\"{o}teman \textit{et al.} \cite{63-goteman2015optimizing} used various layouts in their study and concluded that, except for the rectangular configuration, converters placed away from incoming waves capture less power than those placed closer. According to \cite{19-sharp2018wave}, increasing the number of converters may initially improve array performance, but after a while, additional devices reduce array performance. It is inferred from \cite{14-sharp2015wave}, evaluating the array with optimum power alone is insufficient; however, using a robust optimization method that includes cost properties in the objective function would result in more reliable outcomes. Their proposed GA gives more flexibility in the number of WECs as well as considering multidirectional waves. Valuable studies have been done by Neshat \textit{et al.} \cite{neshat2019new,23-neshat2019adaptive}to optimize WEC positions for maximum power output.  In \cite{neshat2019new}, the answer of 4-buoy layout is mostly aligned and perpendicular to the predominant wave. Likewise, in the 16-buoy layout, the WECs were placed mostly in the feasible area's diagonal. Another publication of Neshat \textit{et al.} \cite{neshat2019adaptive} optimized the layout with a proposed ANSO algorithm to place the WECs. The optimal layout is derived with the sequential placement of the converters where the starting point of placing them is suggested at the bottom right of the area. 
 Table \ref{table:layouts} congregates some of the recent publications in terms of their configurations, number of WECs, type of converter, and determinate the criteria for selecting the master layout.

\begin{figure*}
%[pos=tbp,align=\centering]
\centering
\includegraphics[width=\linewidth]{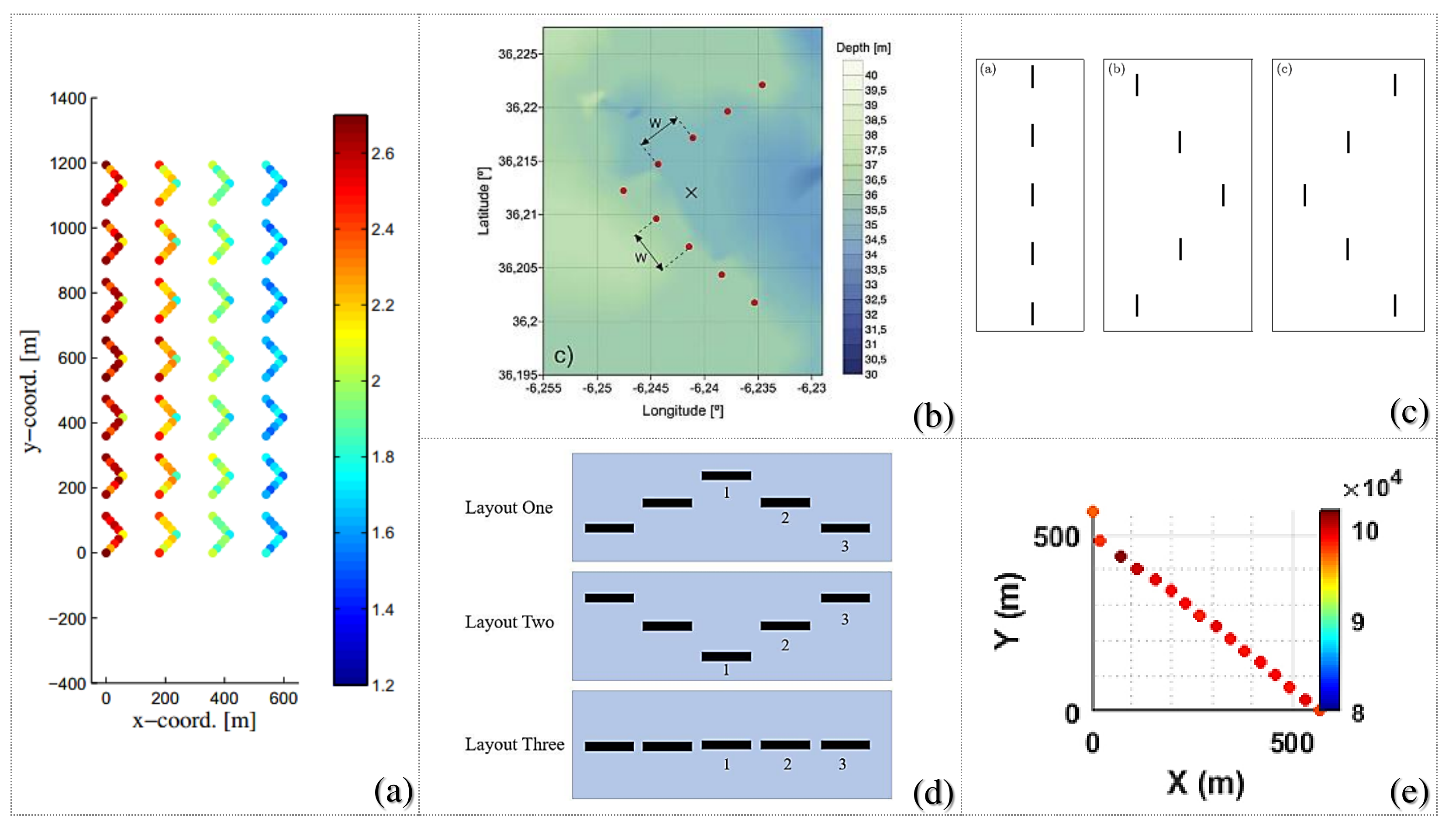}
\caption{Comparative point of view on Arrow (wedge)  as well as linear  pattern of WECs between the recent papers.}
\label{fig:layout}
\end{figure*}

 %%%%%%%%%%%%%%%%%%%%%%%%%%%%%%
\begin{table*}[!htb] 
\centering
\caption{ Layout study in regular and arbitrary patterns in the recent articles.  }
\label{table:layouts}
 \scalebox{0.8}{
\begin{tabular}{|p{16mm}|l|l|l|p{35mm}|p{26mm}|p{24mm}|p{16mm}|}
\hlineB{5}
             
\textbf{Author}   & \textbf{Year} & \textbf{Type of Converter} & \textbf{WECs NO.} & \textbf{Patterns of Layout} & \textbf{Master layout} & \textbf{objective of comparing layout } & \textbf{reference} \\ \hlineB{2}

Hamed Behzad  &  2019 & OWSC & 5 &  arrowhead up and arrowhead down, linear & 	 arrowhead up	& absorbed energy &	\cite{hamedbehzad}
           \\ \hline %
           Ruiz  &  2017 & Surging barges & 25 &	optimized & --& annual power	&	~\cite{ruiz2017layout}
           \\ \hline %
           %%%%%%%%%%%%%%%%%%%%%%%%%%%%%%%
           Giassi  &  2020 & Point absorber& 10-20-50 & optimized &	--	& LCOE &	~\cite{6-giassi2020economical}
           \\ \hline %
           Fang  &  2018 & Point absorber &	 3-5-8 & optimized , line, triangle & --	& absorbed energy &	~\cite{9-fang2018optimization}
           \\ \hline %|?
           Giassi  &  2020 & Point absorber & 6 &	staggered, 2 aligned rows, b-shape & staggered, b-shape &	performance	&	~\cite{11-giassi2020comparison}
           \\ \hline %
           Giassi &  2018  & Point absorber & 4-5-7-9-14	& optimized &	---	& power output &	~\cite{10-giassi2018layout}
           \\ \hline %
           Baltisky  &  2014 & OWSC & 2-3-4-5-6 &	regular polygons & 2 bodies (linear)	& mean annual production &	~\cite{13-balitsky2014control}
           \\ \hline %
           Sharp  &  2015 & Point absorber  & 5 &	optimized &	--&	power and cost &	~\cite{14-sharp2015wave}
           \\ \hline %|?
             Bozzi  &  2017 & point absorber & 4 &	linear, square, rhombus & rhombus, linear	& absorbed energy &	~\cite{15-bozzi2017wave}
           \\ \hline 
           Moarefdoost  &  2016 & Point absorber &	2-3-4-5-6 & optimized & --	& q-factor &	~\cite{16-moarefdoost2017layouts}
           \\ \hline %|?
            Sharp  &  2018 & Point absorber & 5 &	optimized & -- &	power	&	~\cite{50-bharath2018numerical}
            \\ \hline %
           Neshat  &  2019 & Point absorber & 4-16 &	optimized &	-- 	&total power output &	~\cite{20-neshat2020hybrid,21-neshat2019new,23-neshat2019adaptive}
             \\ \hline %
                      Wang  &  2020 & Point absorber & 6 & triangle,	Inverted triangle &	triangle(one buoy faced incident wave) & annual power	&	~\cite{25-wang2020study}
           \\ \hline %
                      Ruiz  &  2018 & overtopping & 9 &	aligned, staggered, arrow &	arrow & absorbed energy	&	~\cite{30-lopez2018methodology,31-lopez2018towards}
           \\ \hline %
                      Bosma  &  2020 & OWC & 5 & staggered, optimized	 & optimized  & average power	&	~\cite{39-bosma2020array}
           \\ \hline %
                      Borgarino  &  2012 & cylinder barge & 9-16-25 &	triangle, square & triangle &	q-factor	&	~\cite{51-borgarino2012impact}
           \\ \hline %
                   Giassi  &  2017 & Point absorber & 4-5-7-9 &	optimized &	--	& q-factor &	~\cite{59-giassi2017parameter}
           
           \\ \hline %
                G\"{o}teman  &  2015 & Point absorber &60-100 &	3-5 C-shaped clusters &	larger circles	& power output &	~\cite{62-goteman2015fast}
           \\ \hline %
                  G\"{o}teman  &  2015 & Point absorber &250  &	random, rectangular, wedge, C-shaped &	Circular (cable length), wedge (power)	& cable length and power &	~\cite{63-goteman2015optimizing}
           \\ \hline %
                  Mcguiness  &  2016 & Point absorber & 5-6-7 &	linear, circular &	circular & q-factor	&	~\cite{67-mcguinness2016hydrodynamic}
           \\ \hline %
Noad  &  2015 & OWSC & 3-5 & linear, staggered, diagonal, bowl-like	 & bowl-like &	absorbed power	&	~\cite{68-noad2015optimisation}
           \\ \hline %
Sharp  &  2017 & OWC & 5 & optimized & --	& generated power &	~\cite{70-sharp2017array}
           \\ \hline %
Wu  &  2016 & Point absorber & 25-50-100 &	optimized  & -- &	q-factor	&	~\cite{78-wu2016fast}
           \\ \hline %
    DeAndres  &  2014 & Point absorber & 2-3-4 &	linear, triangle, square, rhombus &	square	& q-factor &	~\cite{de2014factors}
           \\ \hline %

 \hlineB{5}   
          
\end{tabular}
}
\end{table*}
In comparative point of view, the arrangement of WECs must not be parallel to the wave direction in order to escape the shadowing and masking effects. Two of such layouts are linear and arrow which placed perpendicular to the wave direction. As it can be seen in Figure \ref{fig:layout}, while arrow(wedge) layout is one of the superior positioning showed both face up and down in (b) and (c), the superior arrow layout is the one with its first converter facing the incoming waves, placed at the tip of arrow, similarly to shape b. Also  linear layout perpendicular to wave direction is important to consider and optimisation study of Neshat \cite{neshat2019new} agrees with this layout which is demonstrated in  (a).

%%%%%%%%%%%%%%%%%%%%%%%%%%%%%%%%%%%%%%%%%%

%%%%%%%%%%%%%%%%%%%%%%%%%%%%%%%%%%%%%%%%%%%%

To sum up, more separation between converters and increasing the number of WECs increase the absorbed power. However, after a while, it diminishes the mean harnessed energy. Furthermore, shorter wave period result in more power output. Moreover, dimensions of a WEC needs to satisfy capturing more energy and lower expenditure. Also the master configuration is different in each study. Although in G\"{o}teman review paper \cite{3-goteman2020advances} layout trends are likely to be aligned perpendicular to the direction o predominant wave, it is worth to mention the wedge (arrow) shape and staggered layouts in regular pattern without using optimization methods to return worthful outputs. Finally, using optimization is an alternative choice because it searches all the possible solutions so the results become more viable and secure, and an increase in the number of publications about layout optimizations confirms this. It is concluded that a hybrid algorithm combining the local search method and bio-inspired method introduced by Neshat \textit{et al.} \cite{20-neshat2020hybrid} outperform other heuristic algorithms in terms of accuracy and convergence time.

%%%%%%%%%%%%%%%%%%%%%%%%%%%%%%%%%%%%%%%%%
\subsection{Power Take-off Advancements} \label{subsection2.3}

\begin{figure*}
% [pos=tbp,align=\centering]
\centering
\includegraphics[scale=0.65]{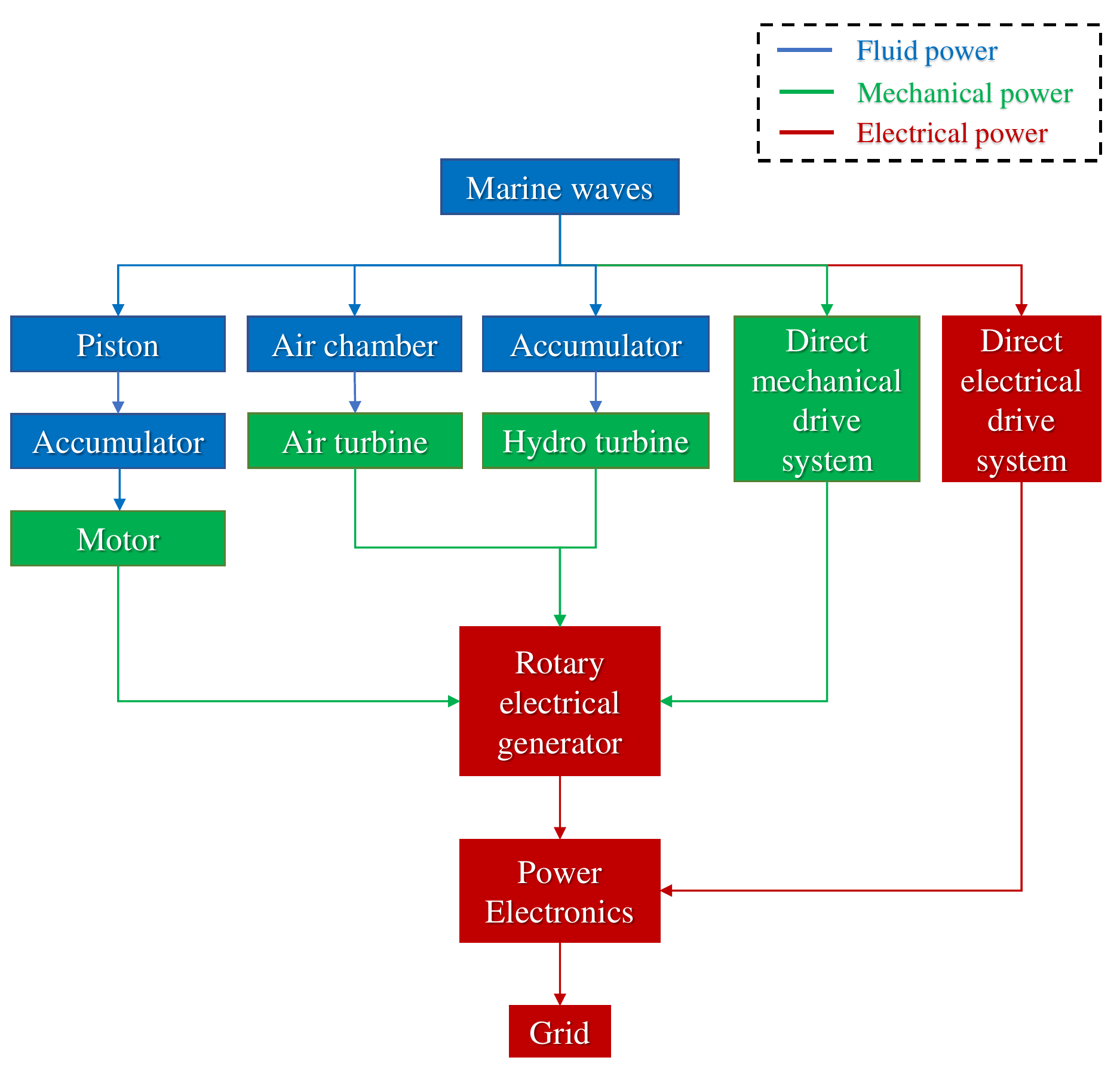}
\captionsetup{justification=justified}
\caption{Different parts and routs of energy conversion in a power take-off system, from wave to grid  (adapted from \cite{pecher2017handbook})}

\label{fig:pto}
\end{figure*}

One of the most vital parts in designing and controlling wave energy converters is the PTO system. The design and control of a PTO system using different strategies can lead to the reduction in WEC's capital cost of energy \cite{pecher2017handbook} which can reduce the LCOE in the long run \cite{6-giassi2020economical}. In fact, optimizing a PTO system for a WEC in a layout has its own challenges regarding two main reasons. First, the irregular fluctuations in water free surface elevation increase the level of uncertainty in the further deterministic analysis. Second, unanticipated changes in WEC's location aquaculture may impose unpredicted forces on the converter, increasing or decreasing its displacement, velocity, or acceleration\cite{babarit2017ocean}. In recent years there have been some research on optimization of PTO settings or control strategies coefficients \cite{41-parrinello2020adaptive, 23-neshat2019adaptive, penalba2017mathematical, ringwood2018wave,amini2021comparative}. In this regard, a variety of power take-off systems has been designed and optimized in the recent literature, as shown in figure \ref{fig:pto}.

\begin{figure*}
% [pos=tbp,align=\centering]
\centering
\includegraphics[clip,width=\linewidth]{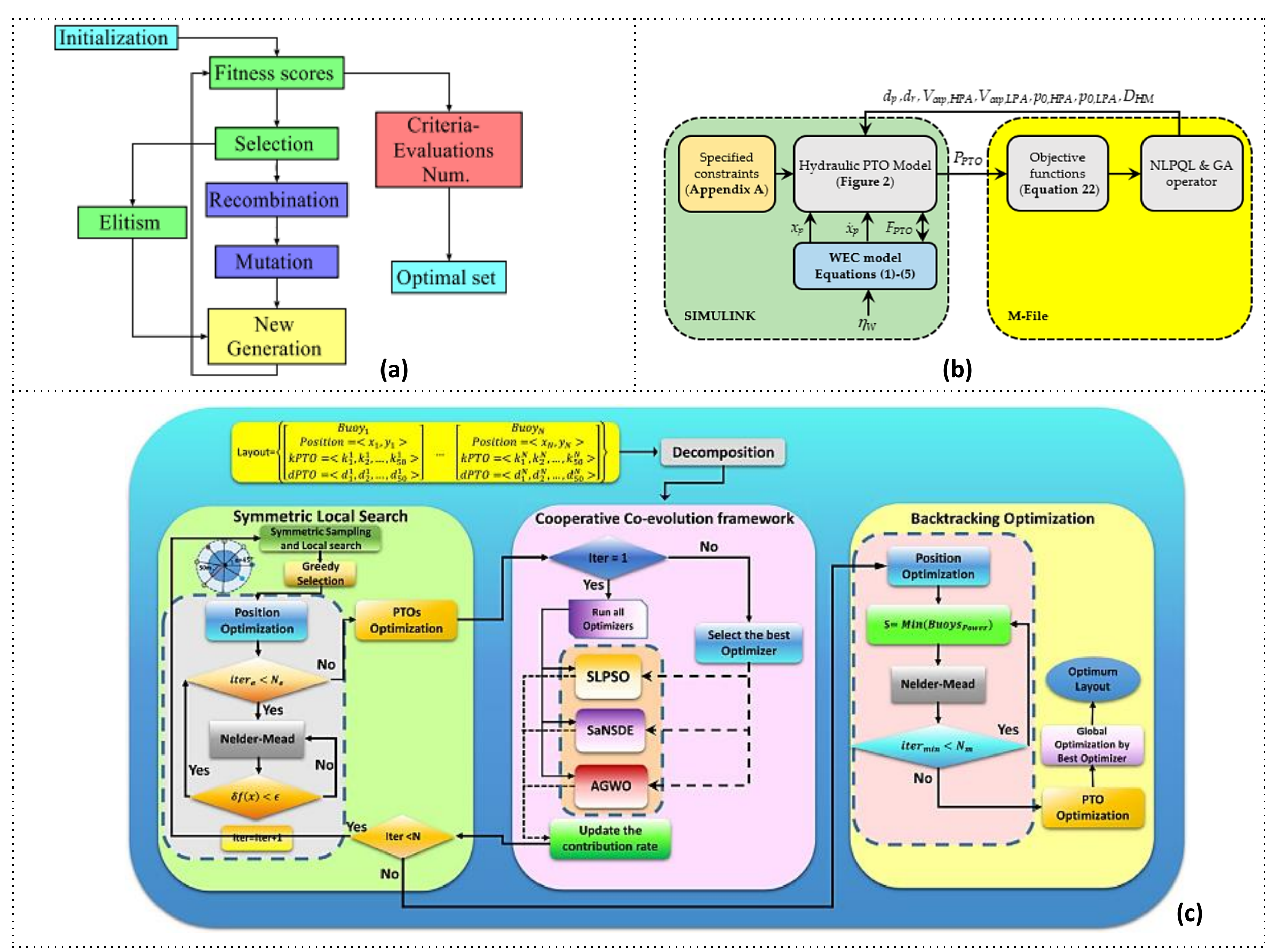}
\caption{Optimization process flowcharts of three recent studies on PTO parameters optimization. (a): optimization flowchart of constructive parameters in EASY software using EAs \cite{76-bonovas2020modelling}. (b): Illustration of optimization model set-up in MATLAB software using NLPQL and GA optimizer \cite{103-jusoh2021estimation}. (c): Outline of the HCCA for PTO hyper-parameters optimization \cite{20-neshat2020hybrid}.}
\label{fig:s3-6}
\end{figure*}

There are approximately 31 active companies in developing direct mechanical approaches in the world so far. For hydro-turbine, hydraulic system, air turbine, and direct electrical systems, these numbers decrease to 21, 13, 13, and 11 active companies, respectively. There are roughly nine other companies developing other power take-off systems that are not classified in this study \cite{79-raju2019heuristic}.\\ 
As a standard approach, regardless of the deployed mechanical equipment, the PTO system of a wave energy layout is mostly designed as a linear spring-damper system, in which generating power is related to Coulomb damping \cite{3-goteman2020advances}. Furthermore, using linear generators can facilitate the direct drive power take-off systems \cite{liu2020survey}. For instance, Vernier hybrid machines \cite{mueller2003modelling}, permanent magnetic synchronous generators \cite{elwood2010design}, switched reluctance linear generators \cite{pan2013voltage}, and  flux-switching permanent magnet linear generators (FSPMLGs) \cite{huang2011novel,huang2013research,huang2014winding} are appropriate to directly convert irregular oscillatory wave motions into unidirectional steady rotation of the generator and produce electricity. Another recent study \cite{dong2021experimental} shows that the wave power is absorbed through the relative motion between the outer and inner cylindrical buoys in a 1:9 scaled model of a two-body heaving WEC. The conclusion is reached that greater relative movement does not imply stronger power capture. This system would benefit from a linear PTO damping system. Figure \ref{fig:s3-6} depicts three recent approaches for PTO parameters optimization regarding the use of Heuristic and meta-heuristic algorithms. 

\begin{figure*}
% [pos=tbp,align=\centering]
\centering
\includegraphics[clip,width=0.8\linewidth]{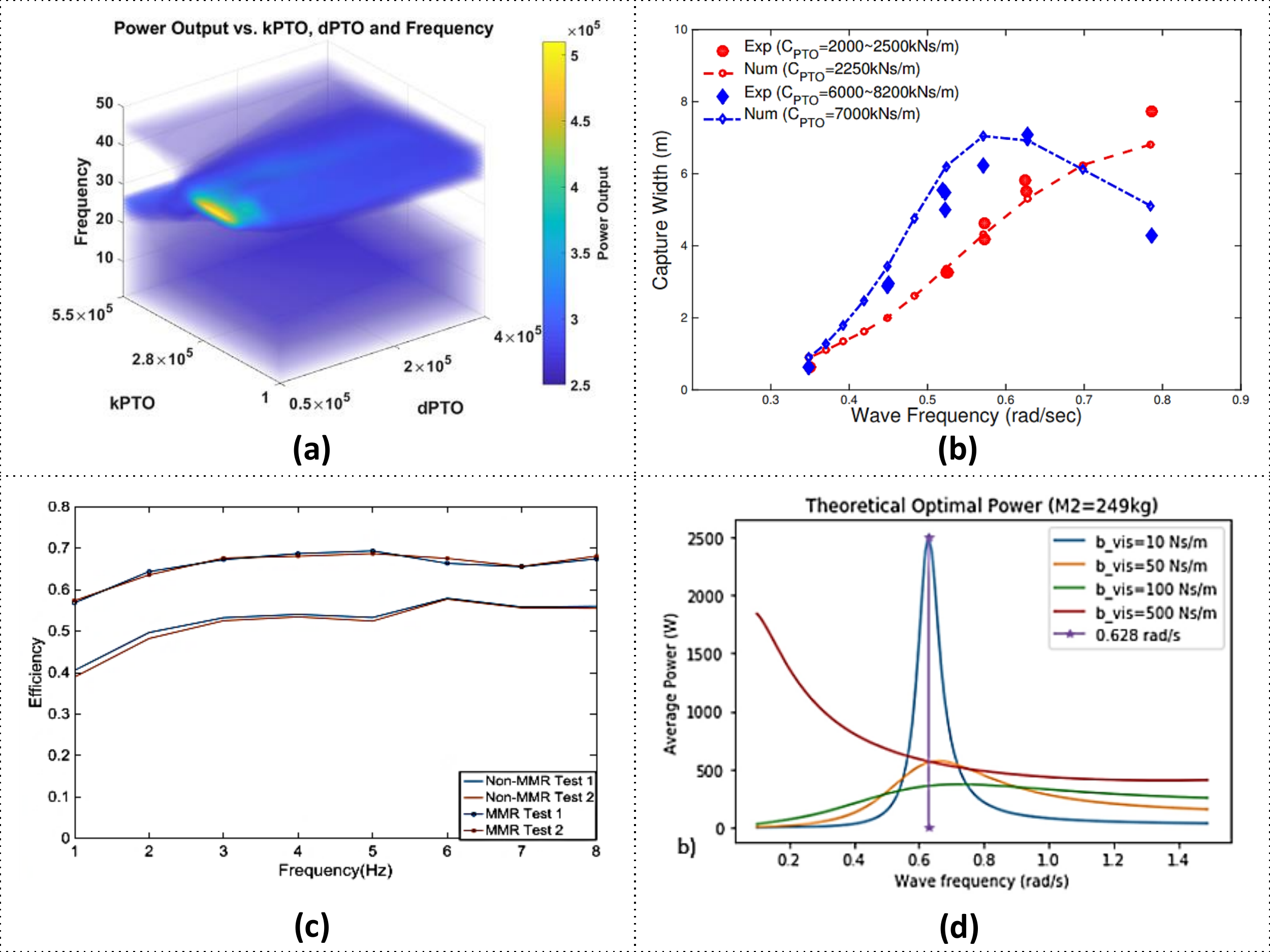}
\captionsetup{justification=centering} 
\caption{Comparing two studies on the effects of PTO coefficient changes on the power output of wave energy converter.(a): 4D view PTO power landscape for a buoy in a layout in the Australian coasts, results from a numerical-optimization study \cite{20-neshat2020hybrid}. (b): Capture width ratio (harnessed power divided by the wave power) over the course of frequency \cite{yu2018numerical}. (c): Lab-testing results over frequency, using MMR-based and non-MMR-based power take-off system\cite{zuo2021efficient}. (d):Average power versus external frequency with optimal $C_{pto}$ and $K_{pto}$ in different linear viscosity damping coefficients \cite{castro2020design}.}
\label{fig:s1-2}
\end{figure*}

As is illustrated in figure \ref{fig:s3-6}, three recently developed solutions for PTO optimizing have been evaluated. First, an optimization studies EAs incorporating EASY software with two specific objective functions: (i) the total investment cost and (ii) the flow rate in the converter's reservoir. The research evaluates WEC's PTO and geometry variables by both single-objective and multi-objective functions shown in figure \ref{fig:s3-6} $(a)$. In fact, the process tries to approach the maximum probable results in terms of total investment cost, flow rate, and annual water storage, according to the selected scenario in the Monterey Bay, California \cite{76-bonovas2020modelling}. Optimization of system's design is performed using general optimization software platform EASY, developed in the NTU Athens \cite{kapsoulis2018evolutionary}.  The research addresses the inefficiency issues by proposing a more advanced piston head design with two functional diameters. This enhanced PTO design is found to increase by 30 percent the annually stored energy in the reservoir \cite{76-bonovas2020modelling}. In the second study flowchart, shown in figure \ref{fig:s3-6} $(b)$, a HPTO model was configured considering the manufacturer's actual hydraulic component parameters. The numerical simulation and optimization process include Non-Linear Programming by Quadratic Lagrangian (NLPQL) and Genetic Algorithm (GA). Although the evaluations are performed for a single rotation-based WEC, the optimized values can be generalized to WEC arrays. Concluding remarks of this study prove the simulation–optimization time with the GA technique to be longer than with the NLPQL process in WEC-modeling problems \cite{103-jusoh2021estimation}. As a much more comprehensive study of PTO hyper-parameters optimization in a WECs array, \cite{20-neshat2020hybrid} suggested a new hybrid cooperative co-evolution algorithm including symmetric LS-NM and a cooperative co-evolution algorithm followed by a backtracking process for optimizing the locations and PTO settings
of WECs, respectively. Figure \ref{fig:s3-6} $(c)$ depicts the proposed hybrid optimization system graphically. After positioning the first buoy in a predetermined position, three optimizers are used to resolve PTO settings for each WEC in layout. Together, the findings confirm the hybrid cooperative system outperforms the others in terms of both runtime and consistency of obtained solutions. As a promising viewpoint, using optimization algorithms to enhance PTO coefficients has been developed recently. Neshat \textit{et al.} \cite{neshat2020hybrid} developed a new optimization framework called HCCA. It comprises a symmetric LS-NM, and a cooperative co-evolution method (CC) with a backtracking strategy. The study's objective was to optimize the positions and PTO settings of 4 and 16 WECs layouts, respectively. Since the PTO stiffness ($K_{pto}$) and damping ($C_{pto}$ or $d_{pto}$) coefficients change over different frequencies, this problem turns into complex multi-directional optimization \cite{20-neshat2020hybrid}. Similarly, Yu \textit{et al.} carried out another numerical study utilizing WEC-Sim solver for the same goal\cite{yu2018numerical}. Although the case studies are different, figure \ref{fig:s1-2} compares both studies and gives us a better understanding of the power output trend proportional to other important variables.

As shown in figure \ref{fig:s1-2} $(a)$, a specific range of frequency with tuned values of PTO coefficients can generate more power output. It can be observed that the overall trend of the power output rises over incoming wave frequency, reaching a maximum amount, and then plummets down to zero at higher frequencies. It certainly depends on the damping coefficient of the PTO system. This study carries out an optimization study to find the optimum value of this coefficient \cite{20-neshat2020hybrid}. Furthermore, by taking other buoys into account, the consequent interactions will increase the complexity of this plot. Although it was not verified with experiments, another experimental study, which is shown in figure \ref{fig:s1-2} (b), can approve the relation between power output and wave frequency. In this study, the power output is represented by capture width in both numerical simulation and wave tank experimental testing \cite{yu2018numerical}. It can be concluded that the more damping coefficient, the more maximum power output could be achieved in lower frequencies. Experimental results in figure \ref{fig:s1-2} $(c)$ confirms the abovementioned idea in the first study. Moreover, the authors proposed an MMR-based PTO system for both prototype (test1 in figure  \ref{fig:s1-2} $c$) and lab testing of PTO (test2 in figure \ref{fig:s1-2} $c$). The results show that MMR-based PTO achieves better efficiency than nonMMR based PTO for the rack pinion design by almost 25 percent at all frequencies \cite{zuo2021efficient}. This path is continued in a further study numerically and experimentally \cite{li2020optimum}. Another similar study suggested a novel configurable electromechanical PTO, which allows setting its parameters for optimal power output according to wave scenarios. In figure \ref{fig:s1-2} $(d)$, the study compares the effects of viscosity damping coefficients on averaged power output over wave frequencies. The resonance condition can be achieved in a specific condition regarding external frequency, and viscosity damping coefficient \cite{castro2020design}.

We described the results of cutting-edge research in PTO setting optimization, which can also be used in WEC arrays. To address further challenges in this regard, more research may be required to fill the gap on optimizing the PTO coefficient over different frequencies. Because the optimization of a set of large WEC power take-off variables is a computationally costly, multi-modal, and large-scale problem, new algorithmic approaches would efficaciously reduce the computational budget. Furthermore, incorporating a smart module to set the PTO spring stiffness continuously in real-time and reducing sliding carriage friction could be another promising approach to enhancing current developments \cite{castro2020design, liu2020survey}. Since this configuration may benefit LCOE, a detailed cost trade-off study would need to be performed \cite{yu2018numerical, balitsky2019analysing}.
%%%%%%%%%%%%%%%%%%%%%%%%%%%%%%%%%%%%%%%%%
\subsection{Geometric Design} \label{subsection2.4}
While increasing the absorbed energy from the WECs will increase profit, it may also increase the costs of the device. Therefore there should be a reasonable balance between these factors to achieve a competitive WEC design \cite{703.clark2019towards}. Since WEC structure is one of the most important factors in minimizing the project costs, to achieve this balance, many studies on WEC hull geometry optimization have been conducted to achieve this balance. A few of these studies are summarized here to provide an overview of general practices in this field \cite{24-garcia2021review}.

First, we take a look at some studies on the geometry optimization of point absorbers. Babarit\textit{ et al.} carried out a multi-objective geometry optimization of the SEAREV device, a floating single body point absorber. Results showed that the largest draughts designs led to the most optimal performances \cite{G52.WREC06}. Another study about a submerged single body point absorber was carried out, in which authors were able to reach significantly higher power output via hull design optimization of the device \cite{G70.1805.08294}. Another study investigated the optimal design of a floating single body point abosorber with 3 different float shapes, using an exhaustive search method. Results showed that despite the fact that the smallest ratios of draft to radius of the floats led to the maximum power output; in 2 of the three investigated cases, it did not led to the most economical design. Furthermore, they found that increasing the mass of the float did not have a meaningful effect on the power output of the WEC \cite{erselcan2020parametric}. Gomes \textit{et al.} optimized the hull design of a two-body point absorber using a GA to maximize the extracted energy, but the obtained dimensions of the WEC were impractical. As a result, they changed their objective function to the ratio of the absorbed energy to immersed volume of the WEC and found a more practical solution. It was found out that increasing the length of the floater has no significant effect on the annual power output \cite{G74.ArtigoConf_13}. In another study, an evaluation of a two-body point absorber was carried out using the Taguchi method in order to increase the power output. The shape of the immersed body was found to be the most important factor in determining power output. Additionally, the buoy diameter and the immersed depth were less critical factors \cite{702.alshami2019}.

The next class of WECs that its hull dimensioning will get a brief review here is OWCs. Bouali \textit{et al.} optimized the geometry of a fixed OWC using a Sequential Procedure, in which they optimized the first parameter, kept it constant, then found the optimal value for the second parameter, etc. They were able to reach an improvement of 7\% in performance \cite{84-bouali2017sequential}.  Likewise, the geometry of a floating OWC was optimized by Gomes \textit{et al.} to maximize the energy output. It was found that the floater’s diameter, the immersed length, and the height of the air chamber can affect the annual power output significantly \cite{88-gomes2012hydrodynamic}.

Next, we will look at two studies optimizing the WEC dimensioning of an OWSC similar to the Oyster. Noad\textit{ et al. }were able to reach an increase of 15\% in the device’s capture factor by decreasing the hinge depth of the device. However, it resulted in big oscillations in the device’s flap, which in turn brought out some inaccuracies in calculations. Therefore they did not use this design in further simulations \cite{G80.flap-opt}. Renzi \textit{et al. }were able to increase the capture factor by approximately 50\% and almost reach a capture factor of 1 \cite{G81.19.Renzi_etal_EWTEC_2017}.

Another class of the WECs is the attenuators. Colby\textit{ et al. }optimized the geometry of an attenuator WEC, which consists of the design of a spar, a forward float, and an aft float to maximize the annual energy extracted.  First, they tried to incorporate a few ballasts and tried to optimize the cuts needed to make the necessary chambers, which led to a 17\% increase in the objective function. Then they attempted to utilize the ability to fill or empty the ballasts, to change their mass parameters, and with a control loop time, they were able to reach a 84\% improvement in performance \cite{G86.10.1.1.380.5526}. Likewise, Wang \textit{et al.} carried out a study to optimize the design of a hinge-barge, which worked as an attenuator, to maximize the energy output using the control-informed geometric design (CIGD) approach, which improved the performance by 22\% \cite{701.LiGuoWang_2019-EWTEC-Geometricoptimizationofahinge-bargewaveenergyconverter}.

The last type of WECs reviewed in this section is the Overtopping WECs. Martins\textit{ et al. }optimized the geometry of a generic Overtopping WEC, using the Exhaustive Search Method to maximize the power output. They tried to change the overtopping ramp slope and the distance between the device and the wave tank bottom. They found out a negative relationship between the ramp slope and the device’s performance \cite{606.j.renene.2017.11.061}. Margheritini \textit{et al.} showed that the geometry of a Overtopping WEC could increase the device’s performance up to 30\% \cite{605.Final2012-TPC-DV01}. More information about WEC geometry optimization studies is presented in Table \ref{table:geometry}.

 %%%%%%%%%%%%%%%%%%%%%%%%%%%%%%
\begin{table*}[!htb] 
\centering
\caption{ Optimisation studies about WEC hull design.  }
\label{table:geometry}
 \scalebox{0.8}{
\begin{tabular}{|l|l|p{18mm}|p{40mm}|p{24mm}|p{48mm}|l|}
\hlineB{5}
             
\textbf{Author}   & \textbf{Year} & \textbf{Type of Converter} & \textbf{optimization Algorithm} & \textbf{Objective Function} & \textbf{optimized Dimensions} & \textbf{Reference} \\ \hlineB{2}

           A. Babarit &  2006  & Point Absorber & Genetic Algorithm	& Absorbed Power, Cost &	the Length, the Beam, the Draught  &	~\cite{G52.WREC06}
           \\ \hline %
           S. Esmaeilzadeh  &  2019 & Point Absorber & Genetic Algorithm &  Power Output & Elongation Coefficients of the Base Shape of the WEC  &	~\cite{G70.1805.08294}
           \\ \hline %
B. Bouali  &  2017 &  OWC & a Sequential Procedure & Hydrodynamic Efficiency & the Immersion Depth, Width of the OWC Chamber Front Wall &	\cite{84-bouali2017sequential}
           \\ \hline %
           R.P.F. Gomes  &  2012 &  OWC & COBYLA, DE  & Energy Absorption & Length and Diameters of the Small and Large Thickness Tubes	& ~\cite{88-gomes2012hydrodynamic}
           \\ \hline %
           %%%%%%%%%%%%%%%%%%%%%%%%%%%%%%%

           I.F. Noad  &  2015 & OWSC &	 -- & Capture Factor & Length, Width of the Flap, Hinge Depth	 &	~\cite{G80.flap-opt}
           \\ \hline %|?
           Emiliano Renzi  &  2017 & OWSC & Genetic Algorithm &	Capture Factor  &	Flap Width, Water Depth, Hinge Height	&	~\cite{G81.19.Renzi_etal_EWTEC_2017}
           \\ \hline %
           Mitch Colby  &  2011 & Attenuator & Evolutionary Algorithm &	 Annual Power Output &  Design of Ballast Chamber Cuts, Weight Distribution &	~\cite{G86.10.1.1.380.5526}
           \\ \hline %
           Liguo Wang  &  2019 & Attenuator  & Exhaustive Search Method &	Extracted Energy & Lengths of the Fore and Aft Barges &	~\cite{701.LiGuoWang_2019-EWTEC-Geometricoptimizationofahinge-bargewaveenergyconverter}
           \\ \hline %|?

 \hlineB{5}   
          
\end{tabular}
}
\end{table*}

Considering the design constraints, dimensions of the WEC hulls were optimized mainly to improve the performance and reduce the hull size, i.e., cost. Results showed that geometry optimization could significantly improve the absorbed energy \cite{G70.1805.08294,88-gomes2012hydrodynamic,G81.19.Renzi_etal_EWTEC_2017,G86.10.1.1.380.5526}. If the PTO control strategy is optimized simultaneously with the geometry using multi-objective optimization, it may result in better solutions \cite{24-garcia2021review}. Although increasing the dimensions of the WECs can potentially result in better performance, it does increase the costs of the operation as well. Thus, there is a trade-off in design between absorbed power and costs that should be addressed. 

\section{Conclusion and remarks}
\label{Section4}
Due to the notable potential of the produced energy from ocean waves in the near future, the development and progression of wave energy technologies are high. WECs technologies need more developments to be commercialized compared to other renewable energies.  In order to achieve the maximum generated power using wave energy converters (WECs), layout and PTO optimization play a significant role. However, optimizing them are challenging because of the complex hydrodynamic interactions among converters.% In this paper, we considered and systematically reviewed the application of various optimization strategies applied to adjust the placement of WECs in a wave farm. Furthermore, the benefits and drawbacks of the concerned optimization methods compared and discussed within different real and synthetic case studies. Finally, we proposed some of the best optimization frameworks which could perform better than other types of techniques to maximise the performance of wave farms.    

In this paper, firstly we discuss about the different classification of converters introduced by \cite{8-drew2009review,babarit2011-2019,Falcao2010wave} and other researchers which we believe that the classification based on working principles together with the hydro-mechanical conversion system is superior one on the grounds of generality as well as considering other distinctive factors. An overview of numerical methods and solvers to unravel the hydrodynamic interactions is given. After that we enumerate the recent articles that used at least one of the methods in their study. We believe that understanding the cause of interaction between WECs and applied forces for the selected converter type must be accurately assumed to start the study, then a low or high fidelity approach should be selected according to the scale and required accuracy of project. After that a survey on optimization problems is done because of the increasing number of algorithms and objective functions studied by the researchers on this field. It is concluded that most of research focus on using GAs with only one objective function. The number of studies related to multi-objective optimization is low. We believe that the use of ANSO, HCCA, and in general, local search method with backtracking optimization can enhance the accuracy. A summary of conclusions about the layout optimization \ref{subsection2.2}, PTO system \ref{subsection2.3}, and geometry optimization \ref{subsection2.4} are as follows.

\begin{itemize}
    \item Layout optimization studies have a wide range of discussing relative factors to reach optimal answer. Two patterns are clearly represented in this paper (linear, arrowhead pattern) to consider as the most repetitive viable results in reviewed studies. Factors such as distance and wave direction are directly affect the layout performance. So we share an overall statement on these factors (increasing or decreasing) in determining the array layout.

\item Using optimization algorithms to improve PTO coefficients and enhance the control strategies has recently been established as a promising viewpoint. Several methods have been used to optimize the PTO systems' coefficient, Including cutting-edge metaheuristic algorithms. Looking at the results of the recent studies, It can be deduced that the higher the damping coefficient, the higher the maximum power output at lower frequencies. This hypothesis is supported by experimental evidence. Furthermore, while PTO system configuration will improve LCOE, a thorough cost-benefit analysis is needed in a variety of PTO systems that are classified in the text. More research is required on active control strategies for PTO system of converters.

\item Studies show that geometry optimization of the WECs can improve the performance notably. Better results can be achieved through geometry optimization in combination with the PTO control strategy. While improving the geometry of the WECs will increase the profit, but performance should be optimized with regards to the increased costs.

\end{itemize}

In the end, we believe that further studies on multi-objective studies where especially consider cost and maximum absorbed energy are more likely to add valuable results and understanding in the future. Furthermore, the absence of numerous studies using CFD rather than BEM as the numerical method is sensed among the published papers. The number of articles using GA is high, so we suggest using other robust metaheuristic algorithms that can return better answers in shorter time. Finally, we wish to see further studies aiming to address the mentioned issues in the future.\\
%------------------------------------------- 

\linespread{1.0}

\section*{Abbreviations}
%\abbreviations{
The following abbreviations are used in this manuscript:\\

\begin{tabular}{c c |c c}

WEC & Wave Energy Converter & LES & Large Eddy Simulation\\
PTO & Power Take-off & SPH & Smoothed Particle Hydrodynamics\\
LCOE & Levelized Cost of Energy & RANS & Reynold-Averaged Navier-Stokes\\
CFD & Computational Fluid Dynamics & BEM & Boundary Element Method\\
GA & Genetic Algorithm & CMA & Covariance Matrix Adaptation\\
DE &  Differential Evolution & CMA-ES & Covariance Matrix Adaptation based Evolutionary Strategy\\
SA & Simulated Annealing Algorithm & NM & Nelder-Mead\\
OWC & Oscillating Water Column & EA & Evolutionary-based Algorithms\\
WES & Wave Energy Scotland & COBYLA & Constrained Optimized by Linear Approximation\\
TWh & Terawatt hour & NLPQL & Non-Linear Programming by Quadratic Lagrangian \\
DOF	& Degree of Freedom & ANSO & Adaptive Neuro-Surrogate Optimization\\
PF & Potential Flow & LS-NM & Local Search plus Nelder-Mead\\
DNS & Direct Numerical Simulation & HCCA & Hybrid Cooperative Co-evolution Algorithm\\

\end{tabular}
%}

\bibliographystyle{unsrt}

\bibliography{Main}

%------------------------------------------- 

\end{document}